\documentclass{aa}
\usepackage{graphicx}
\usepackage{natbib}
\usepackage{multirow}
\usepackage{txfonts}
\usepackage{color}
\usepackage{amssymb}
\usepackage{longtable}
\usepackage{hyperref}

\newcommand{\hii}{H{\sc ii}}
\newcommand{\mia}{IRAS\,19410+2336}

\newcommand{\rxtres}{3\,mm}
\newcommand{\rxun}{1.4\,mm}

\newcommand{\hdco}{\mbox{H$_2$CO}}

\newcommand{\cian}{\mbox{CH$_3$CN}}
\newcommand{\ciana}{\mbox{$\mathrm{CH}_3\mathrm{CN}\left(6_0-5_0\right)$}}
\newcommand{\cianb}{\mbox{$\mathrm{CH}_3\mathrm{CN}\left(6_1-5_1\right)$}}
\newcommand{\cianc}{\mbox{$\mathrm{CH}_3\mathrm{CN}\left(6_2-5_2\right)$}}
\newcommand{\ciand}{\mbox{$\mathrm{CH}_3\mathrm{CN}\left(6_3-5_3\right)$}}
\newcommand{\hdcoa}{\mbox{$\mathrm{H}_2\mathrm{CO}\left(3_{0,3}-2_{0,2}\right)$}
}
\newcommand{\hdcob}{\mbox{$\mathrm{H}_2\mathrm{CO}\left(3_{2,2}-2_{2,1}\right)$}
}
\newcommand{\hdcoc}{\mbox{$\mathrm{H}_2\mathrm{CO}\left(3_{2,1}-2_{2,0}\right)$}
}

\newcommand{\mjybeam}{\mbox{\,mJy\,beam$^{-1}$}}
\newcommand{\colhdos}{\mbox{N$\left(\mathrm{H}_2\right)$}}
\newcommand{\colhdco}{\mbox{N$\left(\mathrm{H}_2\mathrm{CO}\right)$}}
\newcommand{\nhdos}{\mbox{${\rm n\left(H_2\right)}$}}
\newcommand{\tkin}{\mbox{T$_{\rm k}$}}
\newcommand{\abunhdco}{\mbox{$X\left(\mathrm{H}_2\mathrm{CO}\right)$}}
\newcommand{\beuthert}{\citet{beuther2004c}}
\newcommand{\beutherp}{\citep{beuther2004c}}

\renewcommand{\sun}{$_{\odot}$}

\begin{document}
\defcitealias{beuther2004c}{BS04}
   \title{Fragmentation in the Massive Star-Forming Region \mia\thanks{Based on observations carried out with the IRAM Plateau de Bure Interferometer and the IRAM 30m Telescope. IRAM is supported by INSU/CNRS (France), MPG (Germany) and IGN (Spain)}}

	\author{J. A. Rod\'on\inst{1}
			\and
			H. Beuther\inst{2}
			\and
			P. Schilke\inst{3}
			}

	\institute{European Southern Observatory,
				Alonso de C\'ordova 3107, Vitacura, Casilla 19001, Santiago 19, Chile.\\
				\email{jrodon@eso.org}
				\and   				
   				Max-Planck-Institut f\"ur Astronomie,
				K\"onigstuhl 17, 69117 Heidelberg, Germany.
				\and
				I. Physikalisches Institut, Universit\"at zu K\"oln, 
				Z\"ulpicherstr. 77, 50937 K\"oln, Germany.\\
				}

   \date{Received ; accepted }

 
  \abstract
  {The Core Mass Functions of low-mass star-forming regions are found to
resemble the shape of the Initial Mass Function (IMF). A similar result is
observed for the dust clumps in high-mass star forming regions, although at
spatial scales of clusters that do not resolve the substructure that is found in these
massive clumps. The region \mia\ is one exception, having been observed at
spatial scales on the order of $\sim 2500$\,AU, sufficient to resolve the clump
substructure into individual cores.}
   {We investigate the protostellar content of \mia\ at high spatial resolution
at \rxun, determining the temperature structure of the region and deriving its
Core Mass Function.}
   {The massive star-forming region \mia\ was mapped with the PdBI (BCD configurations) at \rxun\ and
\rxtres\ in the continuum and several transitions of formaldehyde (\hdco) and methyl
cyanide (\cian). The \hdco\ transitions were also observed with the IRAM 30\,m Telescope.}
   {We detected 26 continuum sources at \rxun\ with a spatial resolution
down to $\sim 2200$\,AU, several of them with counterparts at NIR and MIR
wavelengths, distributed in two (proto)clusters. With the PdBI \cian\
and PdBI/IRAM 30\,m \hdco\ emission we derived the temperature structure of the region, 
ranging from 35 to 90\,K. Using these
temperatures we calculated the core masses of the detected sources,
ranging from $\sim 0.7$ to $\sim 8\,$M\sun. These masses were
strongly affected by the spatial filtering of the interferometer,
filtering out a common envelope with $\sim 90\%$ of the single-dish flux.
Considering only the detected dense cores, and accounting for binning effects 
as well as cumulative distributions, we derived a Core Mass Function, with a power-law index $\beta=-2.3\pm 0.2$. 
We resolve the Jeans length of the (proto)clusters by one order of magnitude, and only
find little velocity dispersion between the different subsources.}
   { Since we cannot unambiguously differentiate protostellar and prestellar cores, the derived CMF is not prestellar. Furthermore, because of the large fraction of missing flux, we cannot establish a firm link between the CMF and the IMF. This implies that future high-mass CMF studies will require to complement the interferometer continuum data with the short spacing information, a task suitable for ALMA. 
   We note that the method of extracting temperatures using \hdco\ lines becomes less applicable when reaching the dense core scales of the interferometric observations because most of the \hdco\ appears to originate in the  envelope structure.
}


   \maketitle
%

\section{Introduction}
\label{sec-intro}

Our understanding of the structure of the cold, dense interstellar medium (ISM)
in star-forming regions has improved in the last years. In those regions the ISM
exhibits a clumpy, often filamentary structure with density maxima at the sites
of star formation. To represent this structure quantitatively we use the
Core Mass Function (CMF). In this paper we will refer to ``core'' as the small
(diameter ${\rm D}\sim 0.01$\,pc), dense condensation that will form individual
stars or small multiple systems, while with ``clump'' we denote
structures that may form (proto)clusters and may therefore be more
massive and larger than cores. In this sense, cores may be considered as a
subset of clumps.

Sub-mm observations of low-mass star-forming regions such as
Serpens (e.g., \citealt{testi1998}), Orion B (e.g., \citealt{motte2001}), Aquila (e.g., \citealt{konyves2010}) and
$\rho$ Oph (e.g., \citealt{motte1998}), as well as near-infrared extinction maps
(e.g., \citealt{alves2007}), show that their CMFs resemble
the shape and intrinsic mass scale of the stellar initial mass function (IMF;
e.g., \citealt{salpeter1955,kroupa2002}). This suggests that these dense cores
would be the immediate precursors of stars, and that by applying a more or less
constant core-to-star mass conversion efficiency we can obtain the IMF from the
CMF.

In the case of Massive Star-Forming (MSF) regions we have, for example, the
analysis of \citet{reid2006}. They gathered the published masses of the MSF
regions M8 (e.g, \citealt{tothill2002}), M17 (e.g., \citealt{reid2006}),
NGC\,7538 (e.g., \citealt{reid2005}), W43 (e.g., \citealt{motte2003}) and
RCW\,106 (e.g., \citealt{mookerjea2004}), tracing spatial scales of clumps
that correspond to (proto)clusters rather than to individual cores, and tested
the fit of several functional forms for their clump mass functions. They found
that in those cases, the best fit was obtained by a double power law, having a
mean power-law exponent for the high-mass end consistent with the Salpeter IMF.
This would again imply that by an almost one-to-one mass conversion efficiency
the IMF could be obtained from the clump mass function, as in the case for
low-mass star-forming regions. Similar analysis were conducted by e.g.,
\citet{shirley2003,beltran2006}. Furthermore, a theoretical study by \citet{chabrier2010} on the relationship between the CMF and the IMF suggests a tight correlation between the two. Once more, this implies that the IMF would be defined by the CMF in the early stages of evolution.

However, this one-to-one relationship may not hold as, for example, some
clumps must fragment to produce the observed quantity of multiple stellar
systems \citep{goodwin2007}. The relatively large distances ($\gtrsim 2\,$kpc) of
most of the known MSF regions require a spatial resolution of about $1''$ to
resolve the clumps into cores with sizes below $\sim 0.1$\,pc.
That resolution in the (sub)mm regime is only achievable with the
interferometric technique. So far, only a few MSF regions have been observed in
the (sub)mm with spatial resolutions good enough to resolve individual cores
(e.g., \citealt{bontemps2010}, \citealt{fontani2009}, \citealt{rodon2008}, \citealt{rathborne2008},
\citealt{beuther2006c}), and only for one source, \mia, has it been possible to
determine a core mass function \citep{beuther2004c}.

The young MSF region \mia\ is at a distance of $2.16$\,kpc \citep{xu2009} and
has an integrated bolometric luminosity of about $10^4$\,L\sun. It is a very
active star-forming site, with sources detected from X-rays down to radio
wavelengths. It has ${\rm H_2O\ and\ CH_3OH}$ maser emission
\citep{sridharan2002,beuther2002c} and X-ray sources \citep{beuther2002e},
denoting the ongoing formation of intermediate-to-high mass stars. The region
is embedded within a cluster of over 800 components detected at
NIR wavelengths \citep{martin2008,qiu2008}, and it has
a rich and energetic outflow component with multiple outflows detected in CO
\citep{beuther2002b,beuther2003}. \citet{beuther2002a} found that the
large-scale mm emission shows two massive gas clumps roughly aligned in a
north-south direction that splits into several subsources with increasing
spatial resolution (\citealt{beuther2004c}).

With their studies of the mm continuum at high spatial resolution,
\beuthert\ were able to derive the mass function of \mia, resulting in a
Salpeter-like distribution. However, the strongest caveat in the derivation of
that mass function was the fact that a uniform dust temperature was used in the
calculation of the masses. Although they argue that the dust temperature
distribution should not vary strongly, they also warn that changes in the
temperature of the cores would result in a somewhat flattened slope.

We have revisited \mia\ observing the mm continuum at high-spatial resolution
and obtaining molecular-line emission of known temperature tracers, to
determine via several methods a temperature structure for it and
in the end derive a more robust mass function.

\section{Observations}

\subsection{Interferometric}
We observed the two protostellar clusters of \mia\ with the PdBI in the
\textit{B} (Feb-2005), \textit{C} (Dec-2004/Mar-2005) and \textit{D} (Apr-2005) configurations, comprising baselines from 20m to 330m.
This translates into projected baselines ranging from $\sim 7\,k\lambda$ to
$\sim 120\,k\lambda$ at \rxtres\ and from $\sim 15\,k\lambda$ to
$\sim 240\,k\lambda$ at \rxun. The phase centers were set at
$\textrm{RA(J2000)}=19^h43^m10.7^s$;
$\textrm{Dec(J2000)}=23^{\circ}44\arcmin58.4\arcsec$
for the ``Northern'' (proto)cluster and at 
$\textrm{RA(J2000)}=19^h43^m11.2^s$;
$\textrm{Dec(J2000)}=23^{\circ}44\arcmin03.2\arcsec$
for the ``Southern'' (proto)cluster. The continuum was mapped at \rxtres\ and
\rxun.

\begin{table}[h]
	\renewcommand{\arraystretch}{1.2}  
		\caption{Observed molecular transitions and rms of the respective maps.}
		\label{table-lines}
		\centering
		\renewcommand{\footnoterule}{}
		\begin{tabular}{lcccc@{\hspace{15pt}}|c}
			\hline\hline
& $\nu$ & Spectral & E$_{up}$ & \multicolumn{2}{c}{rms\tablefootmark{a}} \\
Transition & (GHz) & Resol. & (K) & \multicolumn{2}{c}{(mJy\,beam$^{-1}$)} \\
\cline{5-6}
 & & (MHz) & & North & South \\
\hline
\ciana\ & 110.383 & 0.18 & 18.5 & 10 & 17\\
\cianb\ & 110.381 & 0.18 & 25.7 & 10 & 17\\
\cianc\ & 110.375 & 0.18 & 47.1 & 10 & 14\\
\ciand\ & 110.364 & 0.18 & 82.9 & 10 & 14\\
\hdcoa\tablefootmark{b} & 218.222 & 0.36 & 21.0 & 29 & 36\\
\hdcob\tablefootmark{b} & 218.476 & 0.36 & 68.1 & 19 & 30\\
\hdcoc\ & 218.760 & 0.36 & 68.1 & 23 & 34\\
			\hline\hline
		\end{tabular}
		\tablefoot{
		\tablefoottext{a}{For a spectral resolution of 0.5\,km\,s$^{-1}$.}
		\tablefoottext{b}{rms values obtained after combining 30\,m and PdBI data.}}
\end{table}

The \rxtres\ receiver was tuned in the upper sideband and the \rxun\ receiver
in the lower sideband. With this spectral setup we observed the \hdco\ and
\cian\ transitions described in Table \ref{table-lines} with a maximum
spectral resolution of 0.5\,km\,s$^{-1}$, adopting a systemic velocity
V$_{\mathrm{LSR}}=22.4$\,km\,s$^{-1}$ \citep{ridge2001,tieftrunk1998}.

The phase and amplitude calibrators were 1923+210 and 2023+336 and the flux
calibrators were 3C273, 2200+420, 1749+096 and 1741-038, adopting for the
calibration the flux values from the SMA flux monitoring of these quasars\footnote{\url{http://sma1.sma.hawaii.edu/callist/callist.html}}. The
data were calibrated with CLIC and then imaged with MAPPING, both part of the
GILDAS package\footnote{\url{http://www.iram.fr/IRAMFR/GILDAS}}. The spectra were processed with CLASS90, also from the GILDAS
package.

After imaging and deconvolution, the resulting synthesized beams for the
continuum data are $1.2''\times0.8''$ at \rxun\ and $2.2''\times1.6''$ at
\rxtres. At the given distance of 2.2\,kpc that means a spatial resolution of
$\sim2200$ and $\sim4200$\,AU, respectively. The continuum data for the Northern
(proto)cluster have rms noise levels $\sigma\sim0.8\mjybeam$ and
$\sigma\sim0.4\mjybeam$ at \rxun\ and \rxtres\ respectively, while in the
southern (proto)cluster the rms noise levels are $\sigma\sim1.0\mjybeam$ and
$\sigma\sim0.4\mjybeam$ at \rxun\ and \rxtres\, respectively.
For the line data, the synthesized beams and noise levels are detailed in Table~\ref{table-lines}.

\subsection{Short spacings}

We obtained IRAM 30\,m observations of \hdcoa\ and \hdcob\ towards \mia\ in
November 2007, taken in the on-the-fly mode with both HERA heterodyne receivers
tuned at 218.222\,GHz. For the backend we used the VESPA correlator, assigning 2
of its spectral bands to each HERA receiver, with a channel spacing of 320 kHz
and a bandwidth of 80 MHz, resulting in a spectral resolution of ~0.5 km/s at
218\,GHz.

The data were processed with CLASS90. The single-dish uv-data were combined with
the interferometric uv-data using MAPPING. After imaging and deconvolution we
obtained a synthesized beam of $1.6\arcsec\times1.0\arcsec$ for the combined
data. The rms levels of the combined data are shown in Table~\ref{table-lines}.

\begin{figure*}[t!]
	\centering
	\includegraphics[angle=-90, width=\textwidth]{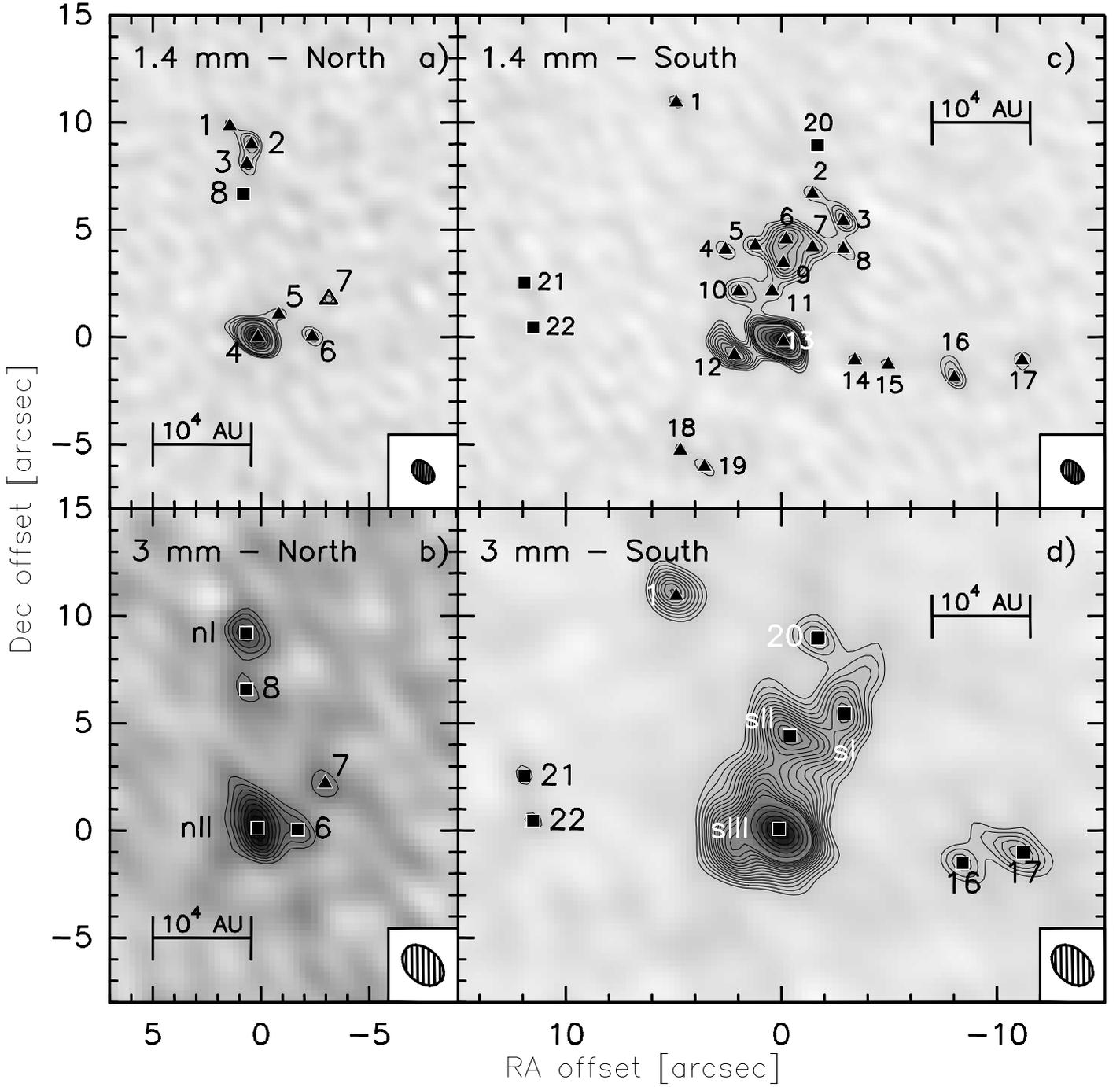}
	\caption{Continuum maps of \mia\ obtained with the PdBI. In the top
row are the \rxun\ maps of the Northern (panel \textit{a}) and Southern (panel \textit{c})
(proto)clusters. Similarly, in the bottom row are the \rxtres\ maps of the
Northern and Southern (proto)clusters in panels \textit{b} and \textit{d} respectively. The
contouring starts at the $4\sigma$ level in all the panels. 
Contours in panel \textit{a} increase in $1\sigma$ steps up to $ 8\sigma $ and in $4\sigma$ steps afterwards,
in panel \textit{b} in $4\sigma$ steps,
in panel \textit{c} in $1\sigma$ steps up to $ 11\sigma $ and in $4\sigma$ steps afterwards, and
in panel \textit{d} in $1\sigma$ steps up to $ 16\sigma $ and in $4\sigma$ steps afterwards. See Table
\ref{table-sources} for the $\sigma$ values.The triangles mark the position of
the sources detected at \rxun\, while the squares are the sources detected at
\rxtres. A square appearing in a \rxun\ map indicates a source that is only
detected at \rxtres. Similarly, a triangle in a \rxtres\ map signals a source
detected at the same position in a \rxun\ map.}
	\label{fig-continuum}
\end{figure*}

\section{Results}

\subsection{Millimetric Continuum}
\label{sec-res-cont}

We established a detection threshold of $4\sigma$ in our \rxun\ continuum maps,
corresponding to $\sim4\mjybeam$, M $\sim 1$\,M\sun\footnote{For
$\tkin\sim40$\,K (see Secs. \ref{sec-res-cont} and \ref{sec-lvg-modeling} for
more details).} and ${\rm \colhdos\sim6\times10^{23}cm^{-2}}$ in the Southern
(proto)cluster, and $\sim3.2\mjybeam$, M $\sim0.8$\,M\sun\footnotemark[1] and
${\rm \colhdos\sim4.5\times10^{23}cm^{-2}}$ in the Northern (proto)cluster. In
the latter we detect 7 sources, while in the former there are 19 sources detected.
Figure \ref{fig-continuum} shows the \rxun\ and \rxtres\ continuum maps for both
(proto)clusters, with the 26 sources detected at \rxun\ marked with triangles
and the two sources only detected at \rxtres\ marked with squares.

The properties of the sources are summarized in Table \ref{table-sources}.
Columns 2 and 3 give their absolute positions, the measured peak flux intensity
and integrated flux density are given in columns 4 and 5 respectively.
For the unresolved sources, their angular size is tipically less than the synthesized beam at the respective wavelength.

\begin{table*}[!t]
	\renewcommand{\arraystretch}{1.2}  
	\renewcommand{\footnoterule}{} 
		\centering
		\caption{Properties of the continuum mm sources in \mia.}
		\label{table-sources}
		\begin{tabular}{lcc|cc|c|ccc|l}
			\hline\hline
\multirow{2}{*}{Source} & R.A. & Dec. & I$_{\nu}$ & S$_{\nu}$ & \tkin\tablefootmark{a} & Mass &
N$\left(\mathrm{H}_2\right)$ & A$_{\textrm{\scriptsize v}}$ & \\
 & (J2000) & (J2000) & (\mjybeam) & (mJy) & (K) & (M\sun) &
($\mathrm{10^{23} cm^{-2}}$) & (10$^3$ mag) & Flags \\ 
			\hline
\multicolumn{10}{c}{\rxun\ with a $1.2''\times0.8''$ beam} \\
			\hline
1-s \ldots & 19 43 11.553 & 23 44 14.15 & 4.8 & \ldots & 35 & 1.5 & 8.2 & 0.9 & $u$ \\
2-s \ldots & 19 43 11.093 & 23 44 09.90 & 5.7 & \ldots & 35 & 1.7 & 9.6 & 1.0 & $u$ \\
3-s \ldots & 19 43 10.989 & 23 44 08.64 & 8.1 & 8.3 & 35 & 2.5 & 14 & 1.5 & \\
4-s \ldots & 19 43 11.387 & 23 44 07.29 & 5.3 & \ldots & 35 & 1.6 & 9.0 & 1.0 & $u$ \\
5-s \ldots & 19 43 11.286 & 23 44 07.49 & 6.2 & \ldots & 35 & 1.9 & 10 & 1.1 & $u$ \\
6-s \ldots & 19 43 11.182 & 23 44 07.78 & 1.1 & 16 & 35 & 4.9 & 19 & 2.0 & \\
7-s \ldots & 19 43 11.093 & 23 44 07.41 & 7.6 & 7.7 & 35 & 2.4 & 13 & 1.4 & \\
8-s \ldots & 19 43 10.989 & 23 44 07.33 & 5.6 & \ldots & 35 & 1.7 & 9.4 & 1.0 & $u$ \\
9-s \ldots & 19 43 11.191 & 23 44 06.68 & 1.0 & 14 & 35 & 4.3 & 17 & 1.8 & \\
10-s \ldots & 19 43 11.342 & 23 44 05.37 & 7.2 & 11 & 35 & 3.4 & 12 & 1.3 & \\
11-s \ldots & 19 43 11.229 & 23 44 05.37 & 5.6 & \ldots & 35 & 1.7 & 9.5 & 1.0 & $u$ \\
12-s \ldots & 19 43 11.357 & 23 44 02.39 & 1.3 & 24 & 45 & 5.5 & 17 & 1.8 & \\
13-s \ldots & 19 43 11.191 & 23 44 03.00 & 4.2 & 75 & 90 & 8.1 & 25 & 2.7 & \\
14-s \ldots & 19 43 10.950 & 23 44 02.15 & 4.6 & \ldots & 35 & 1.4 & 7.8 & 0.8 & $u$ \\
15-s \ldots & 19 43 10.837 & 23 44 01.94 & 4.8 & \ldots & 35 & 1.5 & 8.1 & 0.9 & $u$ \\
16-s \ldots & 19 43 10.614 & 23 44 01.33 & 7.3 & \ldots & 35 & 2.2 & 13 & 1.4 & $u$ \\
17-s \ldots & 19 43 10.385 & 23 44 02.15 & 5.1 & \ldots & 35 & 1.6 & 8.7 & 1.0 & $u$ \\
18-s \ldots & 19 43 11.539 & 23 43 57.94 & 4.6 & \ldots & 35 & 1.4 & 7.8 & 0.8 & $u$ \\
19-s \ldots & 19 43 11.458 & 23 44 57.17 & 5.3 & \ldots & 35 & 1.6 & 9.0 & 1.0 & $u$ \\
1-n \ldots & 19 43 10.801 & 23 45 08.26 & 3.4 & \ldots & 35 & 1.0 & 5.8 & 0.6 & $u$ \\
2-n \ldots & 19 43 10.728 & 23 45 07.41 & 6.1 & \ldots & 35 & 1.9 & 11 & 1.0 & $u$ \\
3-n \ldots & 19 43 10.743 & 23 45 06.51 & 4.5 & \ldots & 35 & 1.4 & 7.6 & 0.8 & $u$ \\
4-n \ldots & 19 43 10.706 & 23 44 58.42 & 2.7 & 32 & 60 & 5.3 & 25 & 2.7 & \\
5-n \ldots & 19 43 10.636 & 23 44 59.48 & 3.6 & \ldots & 50 & 0.7 & 4.1 & 0.4 & $u$ \\
6-n \ldots & 19 43 10.524 & 23 44 58.45 & 4.7 & \ldots & 35 & 1.4 & 7.9 & 0.8 & $u$ \\
7-n \ldots & 19 43 10.469 & 23 45 00.2 & 3.6 & \ldots & 35 & 1.1 & 6.1 & 0.6 & $u$ \\
			\hline
\multicolumn{10}{c}{\rxtres\ with a $2.2''\times1.6''$ beam} \\
			\hline
1-s \ldots & 19 43 11.558 & 23 44 14.41 & 5.1 & \ldots & 40 & 19.9 & 32 & 3.4 & $u$ \\
20-s \ldots & 19 43 11.073 & 23 44 12.23 & 3.1 & \ldots & 40 & 12.1 & 19 & 2.1 & $u$, $n$ \\
c3mm-sI \ldots & 19 43 10.986 & 23 44 08.67 & 4.8 & 6.7 & 40 & 26.2 & 29 & 3.1 & c(2,3,8-s) \\
c3mm-sII  \ldots & 19 43 11.170 & 23 44 07.64 & 6.6 & 13 & 40 & 50.8 & 40 & 4.2 & c(5,6,7,9-s) \\
c3mm-sIII \ldots & 19 43 11.206 & 23 44 03.29 & 17.7 & 48 & 40 & 187.5 & 102 & 10.8 & c(10,11,12,13-s) \\
16-s \ldots & 19 43 10.587 & 23 44 01.71 & 2.9 & \ldots & 40 & 11.2 & 18 & 1.9 & $u$ \\
17-s \ldots & 19 43 10.382 & 23 44 02.20 & 3.5 & 3.9 & 40 & 15.2 & 22 & 2.3 & \\
21-s \ldots & 19 43 12.066 & 23 44 05.76 & 2.0 & \ldots & 40 & 7.9 & 12 & 1.3 & $u$, $n$ \\
22-s \ldots & 19 43 12.037 & 23 44 03.68 & 2.0 & \ldots & 40 & 7.7 & 12 & 1.3 & $u$, $n$ \\
c3mm-nI \ldots & 19 43 10.746 & 23 45 07.63 & 3.2 & \ldots & 40 & 12.7 & 20 & 2.1 & $u$, c(1,2,3-n)\\
8-n \ldots & 19 43 10.746 & 23 45 05.01 & 1.9 & \ldots & 40 & 7.5 & 12 & 1.2 & $u$, $n$\\
c3mm-nII \ldots & 19 43 10.707 & 23 44 58.54 & 5.1 & 7.4 & 40 & 29.0 & 31 & 3.3 & c(4,5-n) \\
6-n \ldots & 19 43 10.573 & 23 44 58.47 & 2.6 & \ldots & 40 & 10.2 & 16 & 1.7 & $u$ \\
7-n \ldots & 19 43 10.480 & 23 45 00.61 & 2.2 & \ldots & 40 & 8.5 & 13 & 1.4 & $u$ \\
			\hline\hline
		\end{tabular}
		\tablefoot{The last column indicates unresolved cores ($u$) and detections at \rxtres\ without a \rxun\ counterpart ($n$). The c(X,Y,Z) flag indicates that the source is the unresolved combination of
sources X, Y and Z. If the source is unresolved, its mass in Column 7 is calculated with the peak intensity value of Column 4.\\
		\tablefoottext{a}{Derived from the \hdco\ emission, the \rxtres\ temperatures are the average in the region (see Sec. \ref{sec-lvg-modeling}). The values for the \rxun\ cores are the same as in Table \ref{table-gauss}, and are repeated here just for reading convenience.}}
\end{table*}

Because the brightness temperature at \rxun\ of the corresponding Planck
function for the strongest source in the region is about 2\,K,
just $\sim 2\%$ of the typical hot core temperatures of $\sim$100\,K,
we can assume that the emission comes from optically thin dust and thus
calculate the masses and column densities with the approach outlined by
\citet{hildebrand1983} and adapted by \citet{beuther2002a,beuther2002erratum}.
We adopted a distance of 2.2\,kpc, and used a grain
emissivity index $\beta = 2$, corresponding to a dust opacity per unit
mass $\kappa_{\rxun}\sim 0.3$ and $\kappa_{\rxtres}\sim
0.08$\,cm$^2$\,g$^{-1}$ for a median grain size
$a = 0.1\,\mu$m, a grain mass density $\rho = 3$\,g\,cm$^{-3}$, and a
gas-to-dust ratio of $186$ \citep{draine2007}. 

The selected value of $\beta = 2$ for the grain emissivity index is the value assumed by \citeauthor{beuther2004c} (\citeyear{beuther2004c}, see also \citealt{beuther2002erratum}), and 
is the typical value found for the ISM dust (e.g., \citealt{hildebrand1983,ossenkopf1994,rodmann2006}). Generally, $ \beta $ is thought to range from 1 to 2. When an IR-to-mm SED is available, as is the case now with \textit{Herschel} results, authors have found $\beta$ values in that range and even higher. Examples are $ \beta\sim1.2$ and $ 1.5 $ in \citet{rathborne2008}, or $ \beta\sim1.7-3.3 $ in \citet{anderson2010} and $ \beta\sim1.0-1.8$ in \citet{rodon2010}. For \mia, there is no data available to directly calculate the dust opacity index. Therefore, in this work we adopted $ \beta=2 $.

In contrast to the bolometer observations of the single dish data (see below), the interferometric continuum data are produced from the line-free part of the spectrum, therefore line contribution to the emission is negligible. Potential contribution from free-free emission is insignificant as well, since the flux at cm wavelengths is on the $ 1 $\,mJy level and only for the central source in the south, with no emission detected from the other sources \citep{beuther2003}. Therefore, we can safely assume that the mm continuum from which the masses are calculated is not severely contaminated by free-free emission.

The calculated masses and H$_2$
column densities are in columns 7 and 8 of Table \ref{table-sources},
respectively. For their calculation we assumed the temperatures shown in
column 11 of Table \ref{table-gauss}. Those temperatures were determined from
the measured \hdco\ line ratios, following the procedure explained and discussed
in Section \ref{sec-deriv-cmf}. The visual extinctions in column 9 were
calculated assuming \mbox{${\rm A_{v}}={\rm N(H_{2}})/0.94\times10^{21}$}
\citep{frerking1982}.

The masses obtained are strongly affected by the spatial filtering
inherent to interferometers. While varying dust properties may account for some of the
discrepancy between the masses calculated at \rxun\ and \rxtres, the
spatial filtering is most likely the main cause.
Comparison with single-dish continuum observations of \mia\ at $1.2$\,mm with the IRAM
30m Telescope \citep{beuther2002a} show that
in our \rxun\ continuum map of the southern (proto)cluster
we are recovering only $\sim6\%$ of the flux estimated with single dish observations.
On the other hand, at \rxtres\ we recover 
a relatively larger amount of the single dish flux, $\sim37\%$. This shows that there is a large fraction of the mass contained in extended structures, that we are filtering out.

The different percentage of flux filtered out in each wavelength band
can be explained in the same way as the large (up to a factor $\sim10$)
difference between the calculated masses at \rxun\ and \rxtres.
Both wavelengths were observed at the same time with the same interferometer
configurations meaning that although the ground baselines are the same, the uv
coverage measured in units of wavelength (k$\lambda$) at \rxtres\ is more
compact than at \rxun, therefore tracing more extended components in the
region.

In the following, we assume that the missing flux is smoothly distributed and affects all cores in the same way.
This assumption comprises one important caveat when deriving a CMF at high-spatial resolution, described in more detail in section \ref{sec-caveats}.

\begin{figure}[ht]
	\centering
	\includegraphics[angle=-90, width=0.5\textwidth]{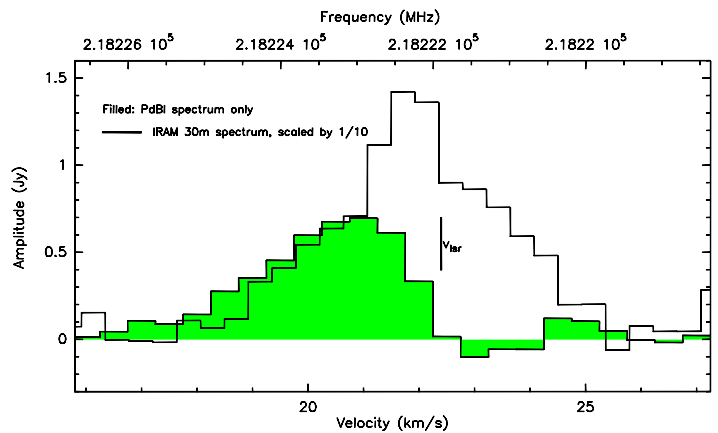}
	\caption{Comparison of the interferometer (PdBI, filled) and
single-dish (IRAM 30m, solid line) \hdcoa\ line emission towards \mia. Clearly
seen is the ``self-absorption''-like feature in the PdBI emission, denoting
the missing flux coming from more extended spatial-scales and filtered out by
the interferometer. The single-dish spectrum has been scaled down for
comparison by a factor 10.}
	\label{fig-shortspacings}
\end{figure}

\begin{figure*}[t!]
	\centering
	\includegraphics[angle=-90, width=\textwidth]{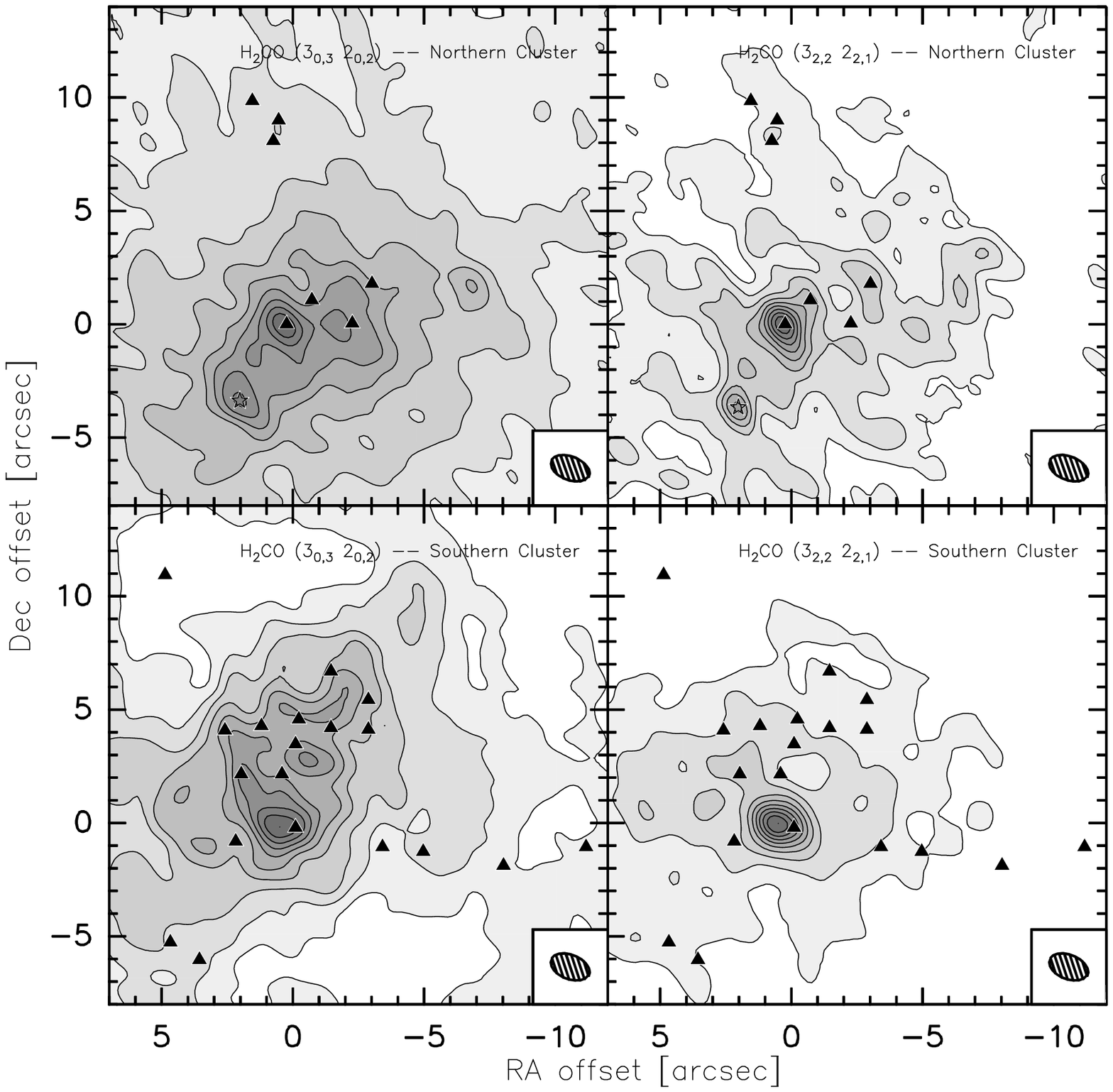}
	\caption{Combined PdBI and IRAM\,30m integrated emission of \hdco\
towards \mia. In the left panels are the \hdcoa\ transition in the Northern
(upper panel) and Southern (lower panel) (proto)clusters, and in the right
panels of the image are the \hdcob\ transition. The contour levels are in 10\%
steps of the peak integrated intensity for each map.
The triangles mark the \rxun\ sources detected, and the $1.6''\times1''$ beam
appears in the lower-right corner of each panel.}
	\label{fig-h2cocombined}
\end{figure*}

\subsection{Formaldehyde}
\label{sec-res-hdco}

With the purpose of estimating the temperature structure of \mia, we
observed three \hdco\ transitions known to be usable as a gas thermometer
(e.g., \citealt{mangum1993}, see Sec. \ref{sec-deriv-cmf}).

Since the PdBI data are clearly affected by missing short spacings (an example of this is shown in Figure \ref{fig-shortspacings}), the region was also observed with the IRAM\,30m telescope. Figure \ref{fig-h2cocombined} shows the integrated intensity maps of the combined PdBI+IRAM 30m data.

Of the three detected \hdco\ lines, \hdcoa\ and \hdcob\ are the ones
with the best signal-to-noise ratio. According to \citet{mangum1993} those two
transitions are enough to determine temperatures, therefore we did not use
the third detected transition, \hdcoc, for the temperature analysis.

Both \hdcoa\ and \hdcob\ have their strongest emission peak towards the
brightest mm source detected in the continuum in both the north and south
(proto)clusters. The secondary peaks in all the maps are in the same spatial
region as the continuum emission, however there is no spatial correlation between the secondary peaks in the \hdco\ and dust emission. This prevents us to obtain a temperature for each individual core.

There is an emission feature seen in the northern cluster in both \hdcoa\ and \hdcob\
transitions $\sim4''$ southeast of the main peak, marked with a star in the upper panels
of Figure \ref{fig-h2cocombined}. The feature is at the rest
velocity, and we do not detect continuum emission neither at \rxun\ nor at \rxtres\ at
that position.

\begin{figure*}[ht]
	\centering
	\includegraphics[angle=-90, width=\textwidth]{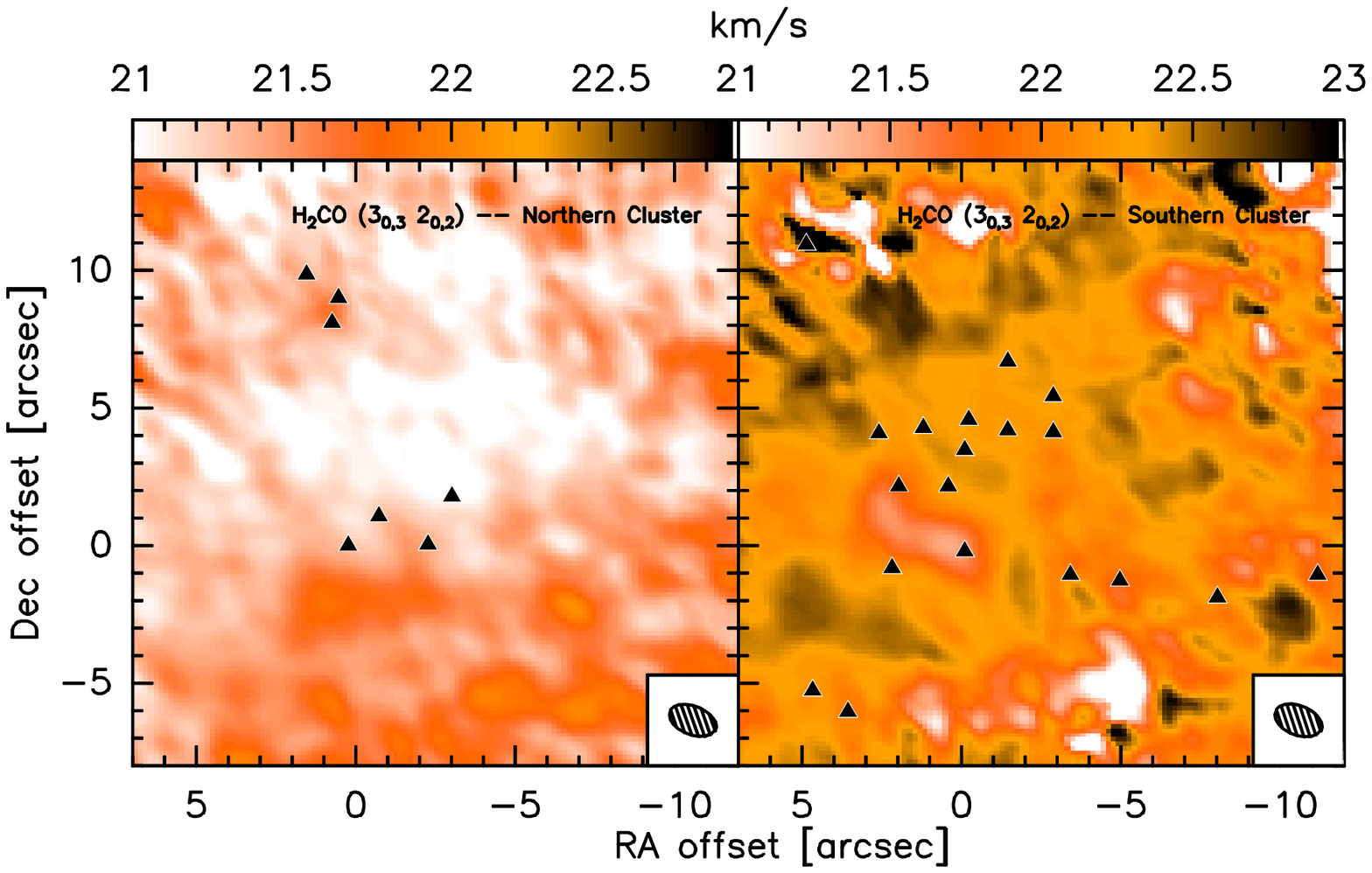}
	\caption{Combined PdBI and IRAM\,30m first moment maps of \hdcoa\
towards \mia\ for the Northern (left panel) and Southern (right panel)
(proto)clusters.
The triangles mark the continuum sources detected, and the beam appears in the
lower-right corner of each panel. There is no signature of a strong velocity
dispersion.}
	\label{fig-h2co-1moment}
\end{figure*}

The first-moment maps of the combined PdBI+30m data (see Fig.
\ref{fig-h2co-1moment}) show only a small velocity dispersion of ${\rm
\sim1\,kms^{-1}}$ in both (proto)clusters for the observed \hdco\ lines.
Similarly, the second-moment maps show no strong variation over the
(proto)clusters.

\begin{table*}[ht]
	\renewcommand{\arraystretch}{1.2}  
		\centering
		\caption{\hdco\ line parameters for the detected cores. }
		\renewcommand{\footnoterule}{}
		\label{table-gauss}
		\begin{tabular}{c|cccc|cccc|cc}
			\hline \hline
 & \multicolumn{4}{|c|}{\hdcoa} & \multicolumn{4}{c|}{\hdcob} & & \\
			\cline{2-9}
Source & $T_{b}$ & $\int T_{b}\,dv$ & $v_{c}$ & Width & $T_{b}$ & $\int T_{b}\,dv$ & $v_{c}$ & Width & Ratio & \tkin \tablefootmark{c} \\
 & (K) & (K\,km\,s$^{-1}$) & (km\,s$^{-1}$) & (km\,s$^{-1}$) & (K) & (K\,km\,s$^{-1}$) & (km\,s$^{-1}$) & (km\,s$^{-1}$) & & (K)\\
			\hline
1-s\tablefootmark{a}	&	\ldots			&	\ldots			&	\ldots			&	\ldots			&	0.4	$\pm$	0.2	&	1	$\pm$	0.4	&	20.4	$\pm$	0.6	&	2.7	$\pm$	1.2	&	\ldots	&	35	\\
2-s	&	14.7	$\pm$	0.5	&	45.7	$\pm$	0.8	&	22.2	$\pm$	0.1	&	3.0	$\pm$	0.1	&	1.7	$\pm$	0.3	&	4.5	$\pm$	0.8	&	21.7	$\pm$	0.2	&	2.5	$\pm$	0.6	&	8.5	&	35	\\
3-s	&	13.6	$\pm$	0.5	&	40.8	$\pm$	0.9	&	22.3	$\pm$	0.1	&	2.8	$\pm$	0.1	&	1.9	$\pm$	0.4	&	3.5	$\pm$	0.4	&	21.3	$\pm$	0.1	&	1.7	$\pm$	0.3	&	7.0	&	35	\\
4-s	&	17.1	$\pm$	0.8	&	51.8	$\pm$	0.8	&	22.3	$\pm$	0.1	&	2.8	$\pm$	0.1	&	3.0	$\pm$	0.3	&	7.4	$\pm$	0.7	&	21.9	$\pm$	0.1	&	2.3	$\pm$	0.3	&	5.8	&	35	\\
5-s	&	17.5	$\pm$	0.7	&	54.4	$\pm$	0.8	&	22.3	$\pm$	0.1	&	2.9	$\pm$	0.1	&	2.7	$\pm$	0.2	&	7.4	$\pm$	0.8	&	21.5	$\pm$	0.1	&	2.6	$\pm$	0.4	&	6.5	&	35	\\
6-s	&	18.0	$\pm$	0.9	&	56.3	$\pm$	1.1	&	22.2	$\pm$	0.1	&	2.9	$\pm$	0.1	&	1.8	$\pm$	0.4	&	6.8	$\pm$	0.8	&	21.4	$\pm$	0.2	&	3.5	$\pm$	0.6	&	9.8	&	35	\\
7-s	&	18.1	$\pm$	0.7	&	54.6	$\pm$	0.8	&	22.2	$\pm$	0.1	&	2.8	$\pm$	0.1	&	2.3	$\pm$	0.2	&	5.9	$\pm$	0.8	&	21.0	$\pm$	0.2	&	2.4	$\pm$	0.4	&	7.8	&	35	\\
8-s	&	12.4	$\pm$	0.4	&	39.0	$\pm$	0.9	&	22.3	$\pm$	0.1	&	3.0	$\pm$	0.1	&	1.8	$\pm$	0.3	&	4.1	$\pm$	0.4	&	20.9	$\pm$	0.1	&	2.2	$\pm$	0.2	&	7.0	&	35	\\
9-s	&	20.5	$\pm$	1.1	&	62.7	$\pm$	0.6	&	22.2	$\pm$	0.1	&	2.9	$\pm$	0.1	&	2.6	$\pm$	0.5	&	7.3	$\pm$	0.5	&	21.6	$\pm$	0.1	&	2.6	$\pm$	0.2	&	7.7	&	35	\\
10-s	&	18.4	$\pm$	1.3	&	61.2	$\pm$	1.3	&	22.3	$\pm$	0.1	&	3.1	$\pm$	0.1	&	2.7	$\pm$	0.4	&	9.3	$\pm$	1.2	&	21.5	$\pm$	0.2	&	3.2	$\pm$	0.5	&	6.8	&	35	\\
11-s	&	19.2	$\pm$	0.7	&	58.3	$\pm$	1.0	&	22.1	$\pm$	0.1	&	2.8	$\pm$	0.1	&	3.0	$\pm$	0.5	&	8.4	$\pm$	0.9	&	21.3	$\pm$	0.1	&	2.6	$\pm$	0.3	&	6.5	&	35	\\
12-s	&	12.8	$\pm$	0.6	&	38.7	$\pm$	0.6	&	22.1	$\pm$	0.1	&	2.8	$\pm$	0.1	&	2.6	$\pm$	0.4	&	8.0	$\pm$	1.1	&	21.3	$\pm$	0.2	&	2.9	$\pm$	0.5	&	4.9	&	45	\\
13-s	&	14.9	$\pm$	1.0	&	83.5	$\pm$	0.8	&	21.6	$\pm$	0.1	&	3.8	$\pm$	0.1	&	4.8	$\pm$	0.5	&	21.7	$\pm$	0.8	&	20.8	$\pm$	0.1	&	3.0	$\pm$	0.1	&	3.1	&	90	\\
14-s	&	5.8	$\pm$	0.2	&	13.3	$\pm$	0.7	&	22.1	$\pm$	0.1	&	2.1	$\pm$	0.1	&	1.7	$\pm$	0.5	&	5.1	$\pm$	0.8	&	21.2	$\pm$	0.2	&	2.9	$\pm$	0.6	&	3.5	&	35	\\
15-s	&	6.5	$\pm$	0.7	&	21.6	$\pm$	1.1	&	22.3	$\pm$	0.1	&	3.1	$\pm$	0.2	&	1.8	$\pm$	0.1	&	\ldots			&	21.4	$\pm$	0.1	&	unresolved 			&	3.6	&	35	\\
16-s	&	3.5	$\pm$	0.2	&	8.6	$\pm$	0.8	&	22.2	$\pm$	0.1	&	2.3	$\pm$	0.3	&	0.8	$\pm$	0.1	&	\ldots			&	21.3	$\pm$	0.2	&	unresolved &	4.2	&	35	\\
17-s	&	4.1	$\pm$	0.3	&	10.8	$\pm$	0.7	&	22.4	$\pm$	0.1	&	2.5	$\pm$	0.2	&	1.8	$\pm$	0.1	&	\ldots			&	22.2	$\pm$	0.1	&	unresolved &	2.3	&	35	\\
18-s	&	8.4	$\pm$	0.7	&	25.3	$\pm$	0.9	&	22.4	$\pm$	0.1	&	2.8	$\pm$	0.1	&	3.1	$\pm$	0.1	&	\ldots			&	21.5	$\pm$	0.1	&	unresolved &	2.7	&	35	\\
19-s	&	8.0	$\pm$	0.6	&	21.6	$\pm$	0.8	&	22.4	$\pm$	0.1	&	2.5	$\pm$	0.1	&	1.0	$\pm$	0.3	&	3.1	$\pm$	0.5	&	22.0	$\pm$	0.3	&	2.8	$\pm$	0.5	&	7.9	&	35	\\
\hline																																					
1-n	&	0.8	$\pm$	0.3	&	3.3	$\pm$	0.7	&	21.3	$\pm$	0.5	&	4.0	$\pm$	1.1	&	0.2	$\pm$	0.1	&	2.3	$\pm$	0.9	&	23.9	$\pm$	1.8	&	9.3	$\pm$	3.5	&	3.4	&	35	\\
2-n	&	1.4	$\pm$	0.2	&	4.2	$\pm$	0.6	&	21.1	$\pm$	0.2	&	3.0	$\pm$	0.6	&	0.8	$\pm$	0.1	&	1.3	$\pm$	0.6	&	20.7	$\pm$	0.3	&	1.6	$\pm$	1.1	&	1.7	&	35	\\
3-n	&	1.4	$\pm$	0.1	&	5.1	$\pm$	0.7	&	21.3	$\pm$	0.2	&	3.4	$\pm$	0.6	&	0.5	$\pm$	0.3	&	2.5	$\pm$	0.9	&	22.8	$\pm$	0.9	&	4.9	$\pm$	2.1	&	2.9	&	35	\\
4-n	&	7.8	$\pm$	0.7	&	35.6	$\pm$	1.1	&	21.0	$\pm$	0.1	&	4.3	$\pm$	0.2	&	2.7	$\pm$	0.4	&	12.2	$\pm$	1.2	&	20.5	$\pm$	0.2	&	4.2	$\pm$	0.5	&	2.9	&	60	\\
5-n	&	5.4	$\pm$	0.2	&	13.6	$\pm$	0.9	&	21.2	$\pm$	0.1	&	2.4	$\pm$	0.2	&	1.8	$\pm$	0.3	&	6.0	$\pm$	1.0	&	20.2	$\pm$	0.2	&	3.2	$\pm$	0.7	&	3.0	&	50	\\
6-n	&	6.8	$\pm$	0.2	&	20.2	$\pm$	0.9	&	21.3	$\pm$	0.1	&	2.8	$\pm$	0.1	&	1.4	$\pm$	0.1	&	3.4	$\pm$	0.8	&	20.7	$\pm$	0.3	&	2.3	$\pm$	0.7	&	4.9	&	35	\\
7-n	&	5.6	$\pm$	0.7	&	16.4	$\pm$	1.3	&	20.6	$\pm$	0.1	&	2.7	$\pm$	0.3	&	1.4	$\pm$	0.3	&	5.9	$\pm$	0.7	&	20.5	$\pm$	0.2	&	4.1	$\pm$	0.6	&	4.2	&	35	\\
			\hline \hline
		\end{tabular}
		\tablefoot{\tablefoottext{a}{No detection in \hdcoa}
		\tablefoottext{b}{Obtained with a $3''$ beam to avoid the optically thick regime.}
		\tablefoottext{c}{From the LVG analysis in section \ref{sec-lvg-modeling}}}
\end{table*}

\subsection{Methyl Cyanide}
\label{sec-res-ch3cn}

\begin{figure}[ht]
	\centering
	\includegraphics[width=0.5\textwidth]{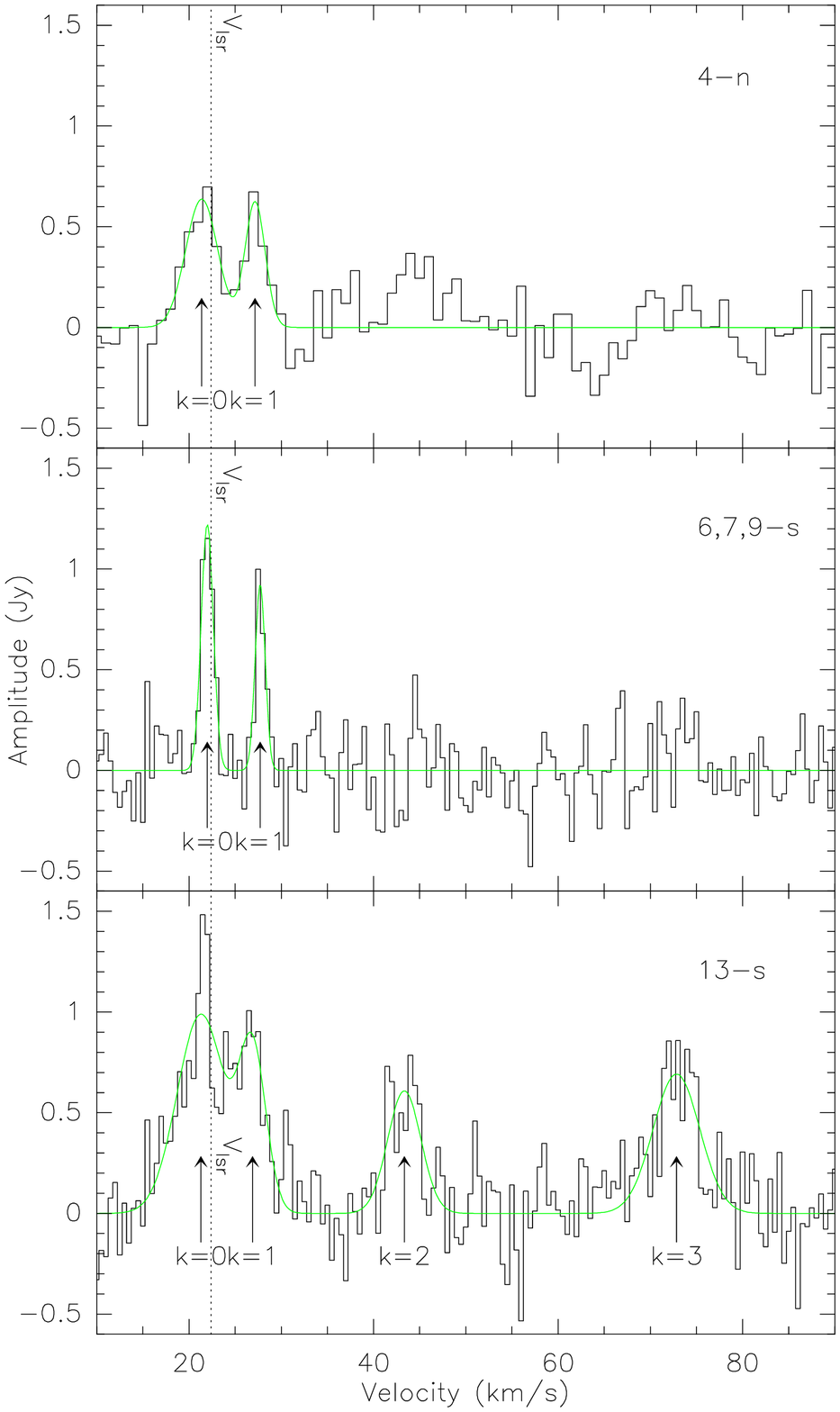}
	\caption{Observed \cian\ spectra towards sources 13-s; 6,7,9-s and 4-n.
The detected $K$-components are marked with arrows in each case, and the rest
velocity with a dotted line. The green solid line is the best Gaussian fit
for the lines, obtained with CLASS, and the resulting line parameters are in
Table \ref{table-gauss-ch3cn}. Only for 13-s were the four $K$-components
detected, indicating a warmer environment.}
	\label{fig-ch3cn}
\end{figure}

We observed the $K=0-3$ components of the $\cian (6_{K}-5_{K})$ ladder,
which are a useful thermometer for the dense gas (e.g., \citealt{loren1984, zhang1998b}). We detected \cian\ emission only
towards sources 13-s, 6/7/9-s and 4-n. Figure \ref{fig-ch3cn} shows the \cian\
spectra toward those three positions.

The $K$=0 and $K$=1 components were detected at the three positions, and towards 13-s we also detect the $K$=2 and $K$=3 components. Since that is the only position at which those transitions are seen, this implies that 13-s is the warmest in the region.
Also there is a tentative detection of the
$K$=2 component towards 4-n.

Unlike with the \hdco\ transitions, \cian\ is a dense gas tracer, and it is barely affected by the missing short-spacings.

\section{Deriving the Core Mass Function}
\label{sec-cmf}

\subsection{Temperature determination}

One of the major caveats of \beuthert\ in deriving a CMF for \mia\
was the assumption of a uniform temperature for both north and south
(proto)clusters when deriving the masses of the cores. In this work,
we try to derive a temperature structure for the two subclusters to obtain more accurate core masses and by that a more reliable CMF.

\begin{figure}[ht]
	\centering
	\includegraphics[angle=-90,width=0.5\textwidth]{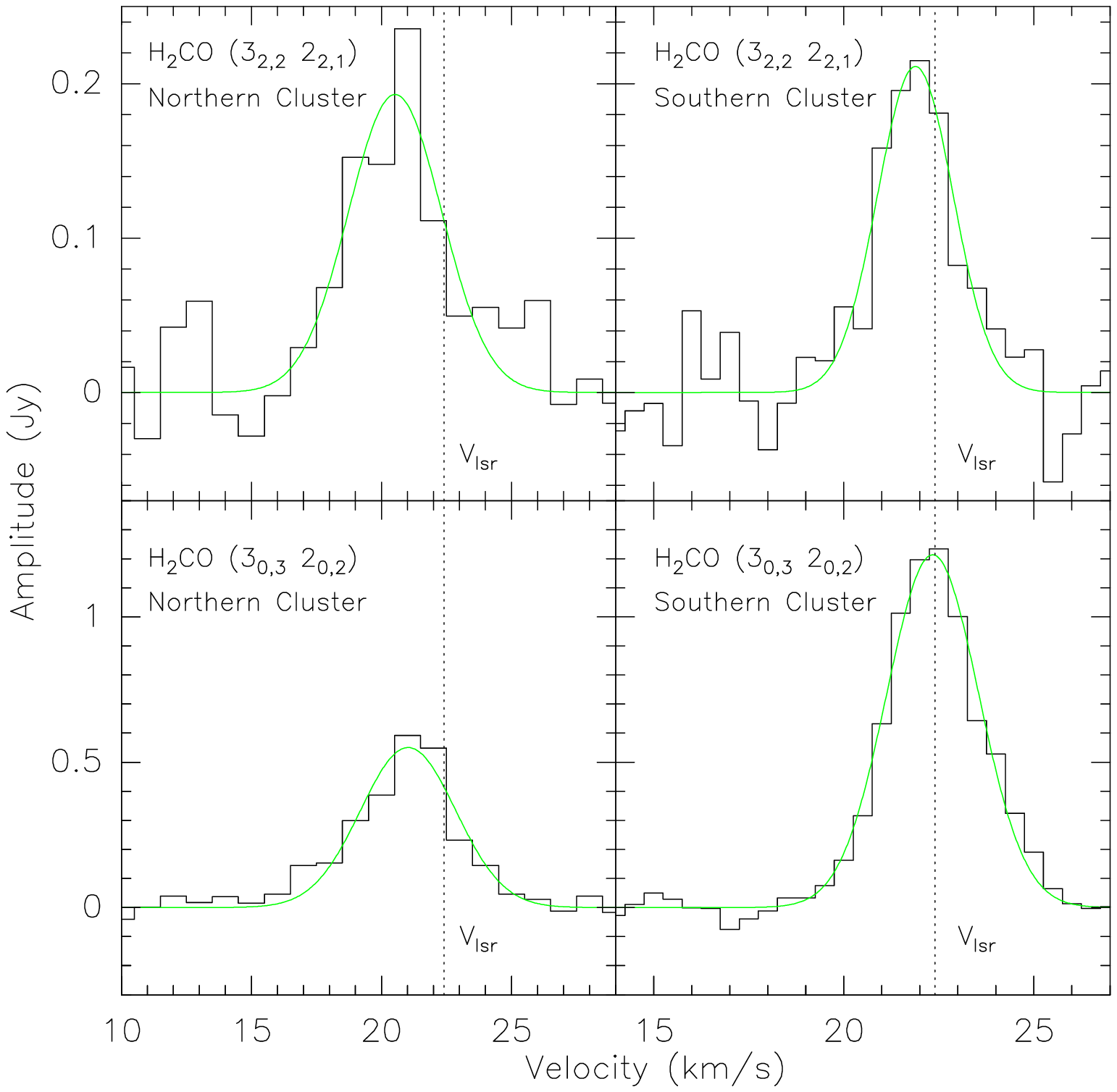}
	\caption{\hdco\ spectra of the main sources in each (proto)cluster of
\mia. In the upper row are the \hdcob\ spectra for sources 4-n (left panel) and
13-s (right panel), and similarly in the bottom row are the \hdcoa\ spectra for
those two sources. A Gaussian profile was fit to each spectrum, and is shown by
a green solid line. The dotted line marks the rest velocity adopted for the
observations.}
	\label{fig-h2co-spectra}
\end{figure}

In this section we describe the process used to determine the temperature of the detected cores. However it has to be taken into account that we are assuming that the kinetic temperatures derived from combined PdBI+30m molecular-line analysis correspond to the dust temperatures of PdBI continuum peaks. As is further discussed in Sec. \ref{sec-caveats} this is not ideal, but is one of the limitations that we still have to face when working with high-spatial resolution mm observations. While combining single-dish and interferometer line data is relatively easy, the vastly different continuum bandpasses of interferometers and single-dish bolometers make the combination of continuum data extremely difficult.

\subsubsection{Applying \hdco\ line ratios}

Formaldehyde $\left(\hdco\right)$, one of the first polyatomic molecules
discovered in space, has proven its usefulness to derive physical
properties of the interstellar gas (e.g.,
\citealt{mangum1993,jansen1994,jansen1995,muehle2007,watanabe2008}). In particular, \hdco\ allows one to estimate
the kinetic temperature and the spatial density in star-forming regions.

Because \hdco\ is almost a symmetric rotor $ (\kappa=-0.96) $, transitions between energy levels with
different $K$ are only collisionally excited. Therefore the comparison between
the level populations from different $K$ components from the same $\Delta J=1$
transition gives a measure of the kinetic temperature of the medium
\citep{mangum1993}.

In this work, we used the line intensity ratios of the \hdcoa\ and \hdcob\
transitions, with energies E$_{up}\sim21$\,K and E$_{up}\sim68$\,K,
respectively. These transitions meet the criteria previously mentioned
and are only $\sim254$\,MHz apart, therefore they could be
observed within the same spectral setup of the PdBI receivers and the HERA
heterodyne receiver. In this way, possible instrumental uncertainties such as
telescope efficiency as a function of wavelength, or receiver and pointing
instabilities, can be disregarded. Hence, one can obtain line ratios that are not
affected by instrumental effects.

Since there is only a poor spatial correlation between the combined \hdco\ and continuum emission, we cannot extract a temperature for each individual core but only a temperature structure of the (proto)clusters as a whole. Only for the brightest continuum cores it was possible to associate the \hdco\ and dust emission. Their spectra are shown in Figure \ref{fig-h2co-spectra}.

Having this temperature information is already an improvement than just assuming a single temperature for all cores. 

\subsubsection{LVG modeling}
\label{sec-lvg-modeling}

We extracted spectra towards the positions of each of the 26 sources detected at \rxun\ and processed them
with CLASS90, the obtained line parameters are listed in Table \ref{table-gauss}.
Once we obtained the $R=\hdcoa/\hdcob$ line ratios, they were compared with
LVG model predictions of the behavior of $R$ as a function of molecular
hydrogen density \nhdos, formaldehyde column density \colhdco, and
kinetic temperature \tkin. 

The H$_{2}$ density could be estimated from our data. With the continuum
emission measured towards the positions of the different sources, we
estimate a beam-averaged H$_{2}$ column density
\colhdos ranging from $\mathrm{\sim5\times10^{23}\,cm^{-2}}$ to
$\mathrm{\sim6\times10^{24}\,cm^{-2}}$ with a mean value of
$\mathrm{\sim10^{24}\,cm^{-2}}$. For this calculation we assumed a priori
a uniform temperature \mbox{$\tkin=46$\,K} \beutherp.

Considering spherical symmetry for the (proto)clusters, at the given distance of
$2.2$\,kpc and if $\theta$ is the angular diameter of the (proto)cluster, the
relationship
\begin{equation}
\mathrm{\nhdos=1.5\times10^{-16}cm^{-3}\left[\frac{\colhdos}{cm^ {-2}}\right]
\left[ \frac{kpc}{d}\right] \left[\frac{arcsec}{\theta}\right]}
\end{equation}
gives an average value $\nhdos\sim 10^7$\,cm$^{-3}$ for each (proto)cluster.
For the calculation we adopted an approximate size of $\theta\sim9''$ for the protoclusters, considering the extension of the main \rxun\ emission. This translates to a size of $ \sim 20000\, $AU at $ 2.2\, $kpc.

In contrast to \nhdos, we cannot directly measure the value of \colhdco\ from
our data. Values for the \hdco\ abundance relative to H$_{2}$ 
are reported in the range of \mbox{$\abunhdco\sim10^{-9} - 10^{-12}$},
for $\nhdos\sim10^{4-6}\,{\rm cm^{-3}}$ (e.g.,
\citealt{wootten1978,mundy1987,carey1998,vandertak2000a}), therefore
calculations of an average \colhdco\ assuming an \abunhdco\ value from the
literature would introduce a high degree of uncertainty.

To constrain the value of \colhdco\ we performed a least-squares
minimization of the \hdcoa\ and \hdcob\ line intensities
as a function of \nhdos\ and \colhdco\ for 13-s, the strongest source in the region.
Assuming T$_{\rm k}=100$\,K (typical for hot cores), we obtain \mbox{$\colhdco\sim 10^{14.5\pm 0.1}{\rm cm^{-2}}$} and \mbox{$\nhdos \sim 10^{6.4\pm1.0}\, {\rm cm^{-3}}$.} as the best solution. This density agrees to within $1\sigma$ with the previously calculated $\nhdos \sim 10^7\, {\rm cm^{-3}}$, while the column density value can be considered as an upper limit since it was derived for the brightest source, with the highest H$_{2}$ column density.

\begin{figure*}[ht]
	\centering
	\includegraphics[angle=-90, width=\textwidth]{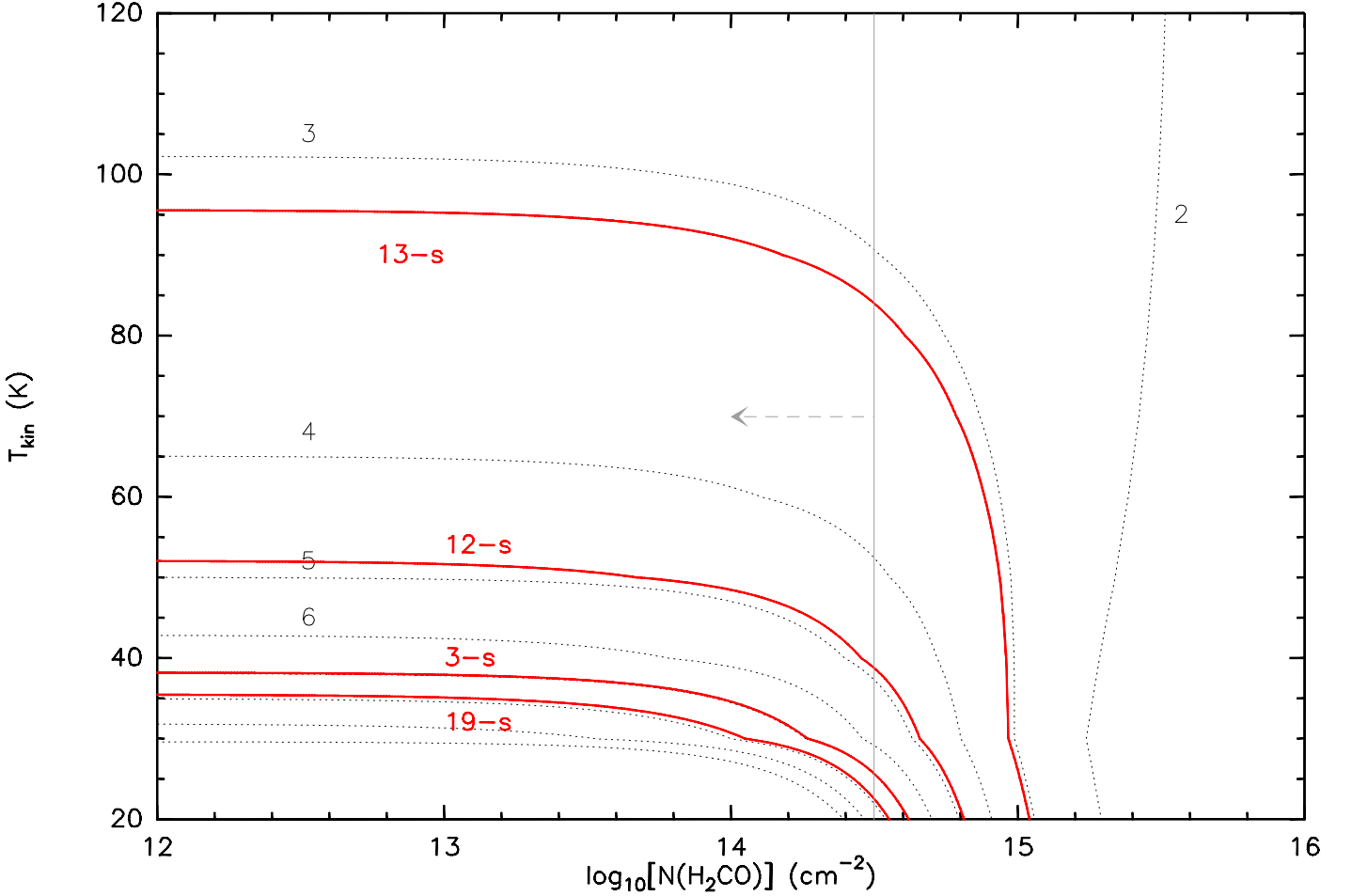}
	\caption{Behavior of the $R=\hdcoa/\hdcob$ integrated intensity ratio. The dotted contours
represent the $R$ values from the LVG model for $\nhdos\sim10^{7}\,{\rm
cm^{-3}}$ (see Sec. \ref{sec-lvg-modeling}) and go from 2 to 10 in unit
steps. The red solid contours are the observed $R$ values, plotted here for
cores 3-s, 12-s, 13-s and 19-s for viewing simplicity. The vertical gray
dotted line is the \colhdco\ upper limit value obtained with the LVG model. The effect
of the optically thin/thick turnover is also seen, which introduces the largest
uncertainty in the \tkin\ estimation.}
	\label{fig-diag}
\end{figure*}

We now compare the LVG model
predictions of the behavior of the ratio $R$ as a function of \colhdco\ and \tkin\ for
$\nhdos\sim 10^7$\,cm$^{-3}$ with the observed $R$ values. Figure \ref{fig-diag}
shows the comparison. The dotted black contours are the modeled $R$ values
from 1 to 10 by $R=1$ steps, while the solid red contours are the observed $R$
for sources 3-s, 12-s, 13-s and 19-s (for viewing simplicity, we only plot a few
sources here). The vertical gray dotted line mark the previously
obtained upper limit value $\log\colhdco\sim14.5$

For $\nhdos\sim 10^7$\,cm$^{-3}$, the calculated upper limit \colhdco\ puts the
cores close to the optically thin/thick regime turnover. This turnover is hinted
at in the behavior of the contour lines in Figure \ref{fig-diag}, and is located
at ${\rm \colhdco\sim10^{15}\,cm^{-2}}$. At higher values the temperature
is no longer sensitive to the column density, which indicates the onset of the
optically thick regime. Another method to (qualitatively) estimate the optical depth is by
comparing the kinetic temperature \tkin\ with the brightness temperature ${\rm
T_b}$. We find that ${\rm T_b}$ is systematically lower than \tkin\ (see Table
\ref{table-gauss}), therefore if we assume beam-filling the medium is optically
thin.

Near to the turnover, the uncertainty in the temperature becomes larger, and in
this case we see that it is approximately $\sim15$\,K. To derive this value we
considered the uncertainty of $\sigma\sim0.2$\,dex in the derivation of
\colhdco. Shifting our upper value for \colhdco\ by that amount, the derived
temperature varies by $\sim15$\,K for 13-s, the brightest source. This can be
seen in Figure \ref{fig-diag}, where it can also be noticed that this effect
is smaller for the sources with lower column density.

As noted before, there is only a poor correlation between the continuum cores and the \hdco\ peaks (except for the main sources).
Therefore, based on their continuum flux and relative location in the
(proto)cluster, we separated the cores into several groups.
For each of these groups an average value of \tkin\ was determined based on the
values obtained for each of the sources within the group. Column 11 of Table
\ref{table-gauss} contains the kinetic temperature values assigned to each
group which are used to determine the masses and \colhdos\ of the cores shown in
Table \ref{table-sources}, and to build the mass distribution to fit the CMF.

The majority of cores appear to be relatively cold, with a temperature of
$\sim35$\,K.
Only cores 12-s, 13-s, 4-n and 5-n have higher temperatures, up to $\sim90$\,K
in the case of 13-s. This indicates that these cores are harboring a
protostar. In the cases of 13-s and 4-n there are
reported NIR counterparts (see Sec. \ref{sec-cont-disc}).
The average temperature value for the whole region is $\sim40\pm15$\,K, in
agreement with the $\tkin\sim46$\,K derived based on IRAS far-infrared
observations \beutherp. Consequently, we find that the mean
\colhdos\ does not deviate significantly from the value adopted at the beginning despite the
changes introduced by the different temperature values (see Table \ref{table-sources}), since that column density was calculated with the IRAS-based temperature.

\subsubsection{The \cian\ k-ladder spectra}

Methyl cyanide (\cian) is a dense gas $\left({\rm n\gtrsim10^{5}\,
cm^{-3}}\right)$ tracer and can also be used as a temperature determinant. As
described previously in Section \ref{sec-res-ch3cn} we only detect \cian\
towards 3 positions, however only for source 13-s we detect enough $k$-components to derive a temperature, and to that source we apply the analysis described in this section. In the other two cases, we only detect the first two $k$-components, indicating that those are cold objects.

\begin{figure}[ht]
	\centering
	\includegraphics[angle=-90,width=0.5\textwidth]{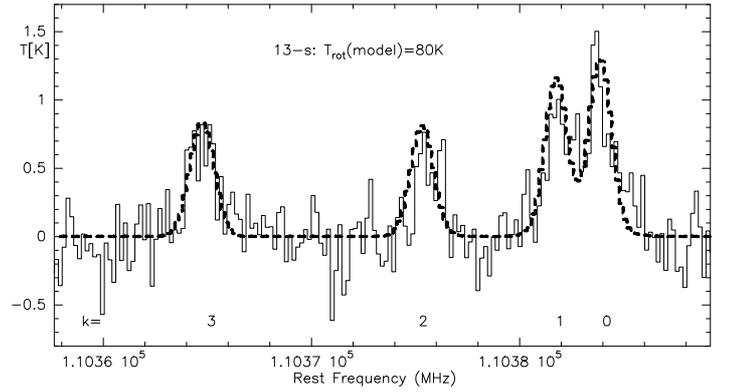}
	\caption{XCLASS (dashed line) fit to the observed \cian\ spectrum
towards source 13-s.}
	\label{fig-ch3cn-xclass}
\end{figure}

Two methods were used. First we compared the \cian\ spectra with
LTE model spectra produced with XCLASS \citep{schilke1999}, a superset of CLASS
of the GILDAS package, obtaining a temperature of $\sim80\pm40$\,K for 13-s (see Fig.
\ref{fig-ch3cn-xclass}). Second, we also derive a
rotational diagram to derive the gas temperature (see e.g.,
\citealt{loren1984,zhang1998b}).

For the rotational diagram, we followed the method outlined in Appendix B of
\citet{zhang1998b}. Basically, we assume LTE and optically thin emission, then
the level populations become directly proportional to the line intensities of
the $k$-components and are translated into a single temperature via the
Boltzmann equation.
With these assumptions, we have the following equation (adapted from eq. [B6] of
\citealt{zhang1998b}) for the relation between the level populations $N_{J,K}$
and the gas temperature $T_{rot}$

\begin{equation}
\label{eq-boltzmann}
	\ln \frac{N_{J,K}}{g_{J,K}} \propto -\frac{E_{J,K}}{k}\frac{1}{T_{rot}}
\end{equation}

\noindent where $g_{J,K}$ and $k$ are the statistical weight of the $(J,K)$ level and the
boltzmann constant, respectively. Studies by \citet{wilner1994} show that
$T_{rot}$ derived with this formulation agrees with those obtained with LVG
calculations.

\begin{figure}[hb]
	\centering
	\includegraphics[angle=-90, width=0.5\textwidth]{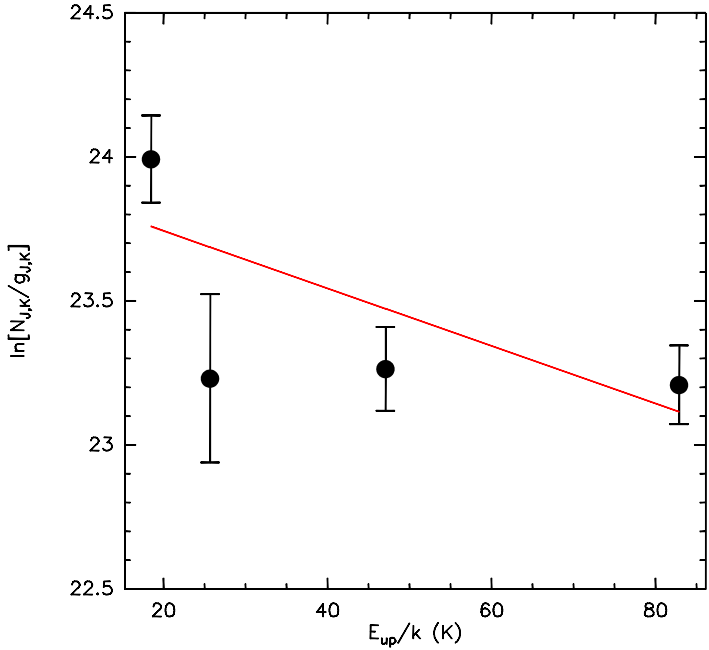}
	\caption{Rotational diagram of the observed $K$-levels of
$\cian(J=6\rightarrow5)$ towards source 13-s. The solid line is the best fit of
Eq. \ref{eq-boltzmann}, corresponding to a rotational temperature of
$100\pm60$\,K, in agreement with the kinetic temperature obtained with the LVG
modeling of the \hdco\ line emission.}
	\label{fig-boltzmann}
\end{figure}

The level population for the upper $k$-level can be obtained from the integrated
intensity of the corresponding line. The line parameters obtained from the
Gaussian fit of the spectra shown in
Figure \ref{fig-ch3cn}, and the calculated $N_{J,K}$ for each level are in Table~\ref{table-gauss-ch3cn}.
The resulting rotational diagram is shown in Figure \ref{fig-boltzmann}, along
with the least-squares fit of all the $k$-components using eq.
(\ref{eq-boltzmann}), obtaining the value $T_{rot}\sim 100\pm60$\,K for 13-s. 

\begin{table*}[ht]
	\renewcommand{\arraystretch}{1.2}
	\centering
	\caption{\cian\ line parameters and LTE column densities.}
	\renewcommand{\footnoterule}{}
	\label{table-gauss-ch3cn}
	\begin{tabular}{cc|cc|cc|cc|cc|cc}
		\hline \hline
\multirow{2}{*}{Source} & \multirow{2}{*}{Line} & $\int T_{b}\,dv$ & $\sigma$ & $v_{c}$ & $\sigma$ & Width & $\sigma$ & $T_b$ & $\sigma$ & $\log N_k$ & $\sigma$ \\
 & & \multicolumn{2}{c|}{(K\,km\,s$^{-1}$)} & \multicolumn{2}{c|}{(km\,s$^{-1}$)} & \multicolumn{2}{c|}{(km\,s$^{-1}$)} &  \multicolumn{2}{c|}{(K)} & \multicolumn{2}{c}{(cm$^{-2}$)} \\
		\hline
\multirow{4}{*}{13-s} & $k=0$ & 6.4 & 1.0 & 21.3 & 0.4 & 6.1 & 1.1 & 1.0 & 0.2 & 12.1 & 0.2 \\
 & $k=1$ & 2.9 & 0.9 & 20.0 & 0.3 & 3.4 & 0.9 & 0.8 & 0.2 & 11.8 & 0.3 \\
 & $k=2$ & 2.7 & 0.4 & 20.2 & 0.3 & 4.2 & 0.7 & 0.6 & 0.2 & 11.8 & 0.1 \\
 & $k=3$ & 4.4 & 0.6 & 19.8 & 0.3 & 6.0 & 1.2 & 0.7 & 0.2 & 12.1 & 0.1 \\
		\hline
\multirow{2}{*}{6,7,9-s} & $k=0$ & 2.0 & 0.2 & 22.0 & 0.1 & 1.6 & 0.2 & 1.2 & 0.1 & 11.6 & 0.1 \\
 & $k=1$ & 1.3 & 0.2 & 20.9 & 0.1 & 1.3 & 0.3 & 0.9 & 0.1 & 11.4 & 0.2 \\
		\hline
\multirow{2}{*}{4-n} & $k=0$ & 2.78 & 0.02 & 21.4 & 0.4 & 4.1 & 0.8 & 0.6 & 0.1 & 11.77 & 0.01 \\
 & $k=1$ & 1.71 & 0.01 & 20.3 & 0.3 & 2.6 & 0.7 & 0.6 & 0.1 & 11.57 & 0.01 \\
		\hline \hline
	\end{tabular}
\end{table*}

From the plot it can be seen that the fit does not represent the
physical
picture well. Explanations of this may be that the assumption of optically thin
emission is not correct, therefore the model applied is not right, or that we
are mapping a hot and small core with a colder extended envelope, thus
presenting two different temperature regimes.

Despite the crude fit, this value agrees very well with the \tkin\ obtained from
the \hdco\ LVG model and also with the value obtained from the XCLASS model.

\subsection{The Differential Core Mass Function}
\label{sec-deriv-cmf}
Combining the data from both (proto)clusters we derive a differential CMF
$\Delta N/\Delta M$, with the number of cores $\Delta N$ per mass bin
$\Delta M$.

One of the strongest caveats (see Sec. \ref{sec-caveats}) we faced when deriving the CMF was the relatively low
number of cores detected. Because of that, a continuous linear binning in mass
was not possible. Instead, we performed a logarithmic binning to better
represent and analyze the data, meaning that the fixed-width mass bins
are defined on a logarithmic axis,
\begin{equation}
	\Delta M  =  \log M_k - \log M_{k-1} = B
\end{equation}
where $B$ is the constant bin width. Therefore, the $k$-th mass is defined as
$M_k = 10^{kB}$. This binning scheme is further discussed in
\citet{maiz-apellaniz2005}.

A priori we do not have a preferred value for $B$, thus we derive a
mass spectrum for different bin widths, with $B$ ranging from 0.001 to 1 in
0.001
steps. Not all the CMFs obtained with this method were fitted. To
obtain meaningful results, we established the following criteria to be satisfied
for a CMF to be fitted:
\begin{itemize}
	\item At most only one bin may contain a single core.
	\item There must be at least four non-empty bins after the
incompleteness threshold.
\end{itemize}

We fit a power-law of the form $\Delta N/\Delta M\propto M^{\beta}$ to the CMFs
satisfying those two criteria, obtaining a $\beta_B$ index value corresponding
to a given $B$. The final value $\beta$ is the weighted mean of all the
$\beta_B$ indices with $\sigma\leq0.4$ and coefficient of
correlation\footnote{This coefficient is defined
as $r^2=1-\frac{SS_{err}}{SS_{tot}}$, where $SS_{err}$ is the sum of squared
errors or residual sum of squares, and $SS_{tot}$ is the total sum of squares. 
By this definition $r^2$ ranges between 0 and 1. A value of 1 means
a perfect fit to the data (e.g, \citealt{draper1998}).} $r^2\geq0.9$.
The different values of $\beta_B$ satisfying these conditions are consistently lower than $\sim-2$, and have a weighted average and formal error $\beta=-2.3\pm 0.2$ (see \citealt{rodon2009phd} for a more detailed description of this method).

We see then that our results for the CMF slope are steeper than the $\sim-1.6$
found for the Clump Mass Functions (e.g., \citealt{stutzki1990,kramer1998}), obtained with CO emission maps for
structures of sizes on the order of $\sim0.1$\,pc. Going to at least one
order of magnitude better in spatial resolution, we find that the slope of the
Mass Function is consistently steeper, indicating further fragmentation at
smaller spatial scales.

\begin{figure}[b!]
	\centering
	\includegraphics[angle=-90, width=0.5\textwidth]{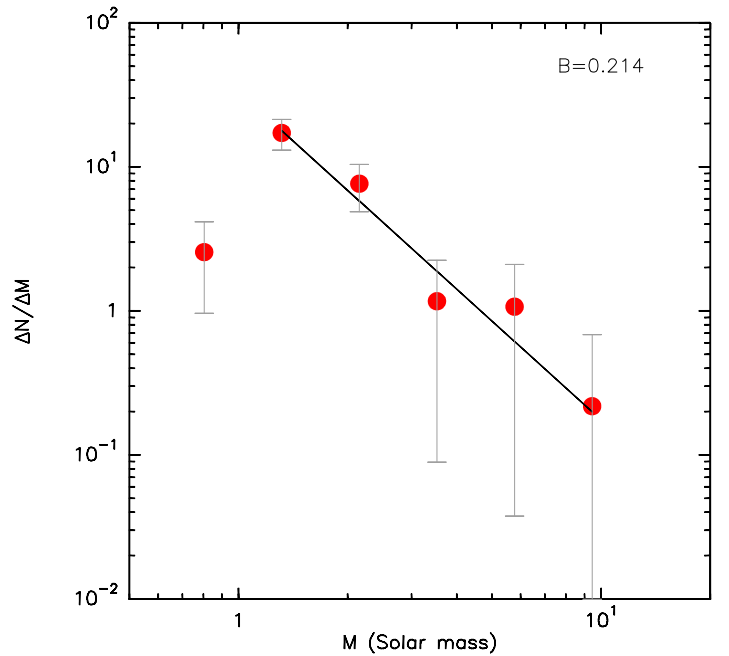}
	\caption{Example of the CMFs obtained, corresponding to $B=0.214$ for
core masses above $1.4$\,M\sun. The solid line is the best fit of eq.
(\ref{eq-cmf-diff}), with a power-law index $\beta=-2.3\pm0.2$.}
	\label{fig-cmf}
\end{figure}

For reference, Figure \ref{fig-cmf} shows an example of the CMFs we obtained,
in this case corresponding to $B=0.214$ and having a power-law index
$\beta=2.3\pm0.2$. A turnover in the distribution at the bin centered at
$\sim1.4$\,M\sun\ can be seen, containing masses above
$\sim1$\,M\sun. It appears in all the derived CMFs at about
the same position, and since it matches our detection threshold, it is likely caused by the low-mass incompleteness of our sample rather
than being a physical feature. In all of the cases, we fit our mass
distributions for masses higher than the turnover.

The incompleteness comes both from both the sensitivity limit, as well as the crowding in the region. The sensitivity limit has been addressed in Section \ref{sec-res-cont}, giving a lower completeness limit of $ \sim1$\,M\sun. We find that the effect of crowding cannot be properly quantified with the given data. Nevertheless, in (sub)mm studies the turnover in the mass distribution and the incompleteness of the sample tend to occur at the same position (see \citealt{kirk2006}, and references therein), therefore we assume that our sample is completeness-limited for masses below $ \sim1 $\,M\sun.

Although we are filtering out considerable fractions of flux (section \ref{sec-res-cont}) on large spatial scales, this filtering likely affects all observations in the
same way. Although we cannot draw conclusions about the absolute masses of
the cores it is expected that their relative masses are correct.

\subsubsection{The Cumulative Mass Function}
\label{sec-cumul-cmf}

Using the differential CMF in a low-number sample has the disadvantage that it
is sensitive to the arbitrariness of binning, as we have shown in the previous
section and led to the analysis described there. On the other hand, a
cumulative CMF is free from any problems associated with the binning, and is
more suitable for low-number samples.

With our treatment of the CMF described in the previous section, we have taken
into account the arbitrariness of the binning. However, for comparison and
further analysis we also derived a cumulative CMF for our sample. Since our
differential CMF above $\sim1$\,M\sun\ can be fit by a single power-law of
the form

\begin{equation}
\label{eq-cmf-diff}
	\frac{\Delta N}{\Delta M}\propto M^{\beta}
\end{equation}

\noindent then the cumulative CMF is

\begin{equation}
\label{eq-cmf-wrong}
	N(>M)\propto-\frac{1}{1+\beta}M^{1+\beta}
\end{equation}

\noindent for $\beta<-1$. According to \citet{reid2006}, when using the cumulative CMF
one should take into account the upper mass cutoff $M_{max}$ of the sample,
whether it is a real cutoff or a result of the finite sampling. In that case,
the cumulative CMF takes the form

\begin{equation}
\label{eq-cmf}
	N(>M) \propto \left\{\begin{array}{lr}
\frac{1}{1+\beta}\left(M_{max}^{1+\beta}-M^{1+\beta}\right), & M<M_{max} \\
0, & M\geq M_{max}
\end{array} \right.
\end{equation}

\noindent for $\beta<-1$. However, they also state that a ``steep power law''
($\beta=-2.5$) will overwhelm the effect that the upper mass cutoff introduces
in the fitting. From the differential CMF we see that we are close to that
``steep power law'' limit, so we adopt both analytical forms given in equations
(\ref{eq-cmf-wrong}) and (\ref{eq-cmf}) for the cumulative CMF.

\begin{figure}[h]
	\centering
	\includegraphics[angle=-90, width=0.5\textwidth]{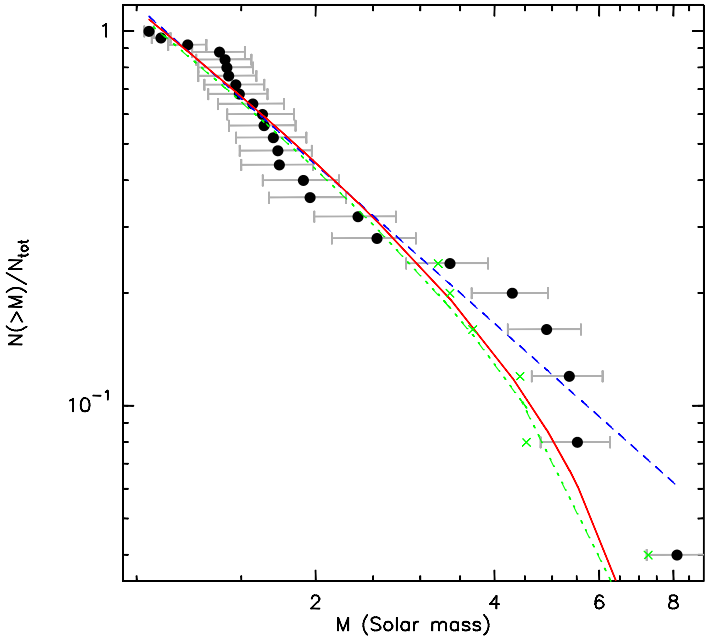}
	\caption{Cumulative CMF of \mia\ for core masses above $1$\,M\sun.
The solid red line and the dashed blue line represent the best fit of equations
(\ref{eq-cmf}) with a power-law index $\beta=-2.2\pm0.1$ and
(\ref{eq-cmf-wrong}) with $\beta=-2.4\pm0.1$, respectively.
The green crosses
represent a slight increase of 10\,K in the higher-mass end, and the green
dash-dotted line is the corresponding new fit of equation (\ref{eq-cmf}) (see Sec.
\ref{sec-disc-beuther}). Notice the flattening of the ``bump'' in the
distribution, while the slope of the fit practically does not change.
In all cases we have normalized $N(>M)$ by the total number of cores
$N_{tot}$.}
	\label{fig-cum-cmf}
\end{figure}

Figure \ref{fig-cum-cmf} shows the cumulative CMF for masses above 1\,M\sun. As
stated before we fit both the analytical expressions given in equations
(\ref{eq-cmf-wrong}) and (\ref{eq-cmf}). The results are $\beta=-2.4\pm0.1$ and
$\beta=-2.2\pm0.1$, respectively, with the $\sigma$ values resulting from the
fitting algorithm. These results are comparable, showing that for a steep power
law the inclusion of $M_{max}$ in the definition of the cumulative CMF
does not affect the result significantly \citep{reid2006}. Since both results
are similar, we adopt their average value $\beta=-2.3\pm0.2$.

The cumulative CMF shows that the lower-mass objects dominate the fit.
This is already visible in the differential CMF, but is clearer in the
cumulative CMF. This is because the lower-mass bins are more populated. We see
that of the 25 cores with masses above $1$\,M\sun\ only 6 are more massive than
$2.5$\,M\sun, therefore the lower-mass bins have more statistical weight in the
fitting.

There is also an inflection in the distribution starting at
$\sim4$\,M\sun. It is not clear if it is a real physical feature of the
distribution or a product of the uncertainty of its derivation. We discuss this
in further detail in Section \ref{sec-disc-cmf}.

\section{Discussion}

\subsection{Comparison with \beuthert}
\label{sec-disc-beuther}

\beuthert\ detected 24 mm sources at $1.3$\,mm, 12 in each northern and
southern cluster, with a detection threshold of $3\sigma\sim 9\mjybeam$. Having
better sensitivity, we adopted a detection threshold of $4\sigma\sim4\mjybeam$
at \rxun\ finding 26 cores, 19 in the Southern cluster and 7 in the Northern
(proto)cluster. Despite the slight difference in the total number, the general
\rxun\ high-resolution cluster structure shown in \beuthert\ is recovered.

This highlights the general mapping difficulties with interferometers
having only a small number of antennas close to the detection limit. For
example, contrary to expectations, formal $3\sigma$ limits in such maps are not
as reliable as one may expect. Therefore, here we raised the threshold to
$4\sigma$. ALMA, with its many antennas, will overcome such problems.

From the analysis described in Section \ref{sec-deriv-cmf}, we obtain a
power-law CMF for \mia\ with an index \mbox{$\beta=-2.3\pm 0.2$}, in
agreement with the previous result of \beuthert.
The new result is more reliable than the previous one since with the new data presented here for \mia\ we were able to estimate a temperature structure for the (proto)clusters. Despite the limitations described in sections \ref{sec-lvg-modeling} and \ref{sec-caveats}, that allow us to obtain a more accurate value for the core masses and thus avoid the previous caveat of assigning a uniform temperature to the whole region.

Nevertheless, this similarity between the CMF indices would imply that there is not a strong dependence with the temperature structure. If we calculate the cumulative CMF of our region, but assuming a uniform temperature $ T=35$\,K, the new index is $ \beta=-2.3\pm0.1 $. In the same way, if $ T=40$\,K then $ \beta=-2.6\pm0.2 $. Both values are indistinguishable within the errors from our main result.
This feature is in part due to temperature regime. The functional dependence of the mass with the temperature implies that if we were to vary the latter from $ 10 $\,K to $ 20 $\,K, the mass would change by a about a factor 3. But, going to higher temperatures, the variation in mass is less pronounced. For example, a change from $ 40 $\,K to $ 50 $\,K implies a change in mass of less than a factor 1.5, decreasing with increasing temperatures. Our other uncertainties in the temperature and mass calculations have more weight than this effect.
Nevertheless, this suggests that a detailed temperature structure plays a more significant role in the determination of the mass functions of true pre-stellar regions, because of their intrinsic lower temperatures. The calculation of the mass function of a star-forming region before its members become stars is key in determining how the parental cloud further fragments, and wether the IMF is set at the fragmentation stage or is a later result of further dynamical processes (e.g., competitive accretion, mergers, etc.).

The calculation of the power-law index was done taking into
account the arbitrariness of the binning at the moment of deriving a
differential CMF. Also, the cumulative CMF was taken into account in the
analysis.

In the case of the cumulative CMF, the result obtained from it matches the
result from the differential CMF. Although the former is better suited
than the latter for the analysis of a low-number sample, in this case
we observe that the fit to the cumulative CMF represents the lower-mass end
of the distribution well but not the higher-mass end. However, a slight 10\,K
increase in temperature in the higher-mass end, which is within the uncertainty in the temperature determination, flattens the inflection
seen starting at $\sim4$\,M\sun, obtaining also a better fitting for the distribution
while not introducing substantial modifications to the fitted parameters.

\subsection{Continuum Sources}
\label{sec-cont-disc}

In the Southern (proto)cluster, there are NIR and MIR counterparts detected for
several of our mm sources.
\citet{martin2008} and \citet{qiu2008} detect in the position of 13-s a
bright source in the $K_s$ filter and the $3.6\mu m, 4.5\mu m$, $5.8\mu m$ and
$8.0\mu m$ Spitzer/IRAC bands. \citet{martin2008} suggest that the detected NIR
and MIR emission is either leaking through an outflow-created cavity (see
\citealt{beuther2003}), or that the cavity itself is radiating the emission,
based on their estimation that the visual extinction at that position should be
very large and thus should not have any detectable NIR or even MIR emission. We
confirm their estimation, finding that the visual extinction for 13-s is
$A_v\sim2700$ (see Table \ref{table-sources}).
Source 13-s is in an early stage of evolution, according to its NIR excess and
the presence of H$_2$O and Class \textsc{ii} CH$_3$OH masers
\citep{beuther2002c}, however the detection of a VLA $3.6$\,cm source at its
position \citep{sridharan2002}, suggest the presence of a recently ignited
protostar that has already formed an Ultracompact or Hypercompact \hii\
region, the detected radio emission being consistent with an ionizing B2 V star
\citep{martin2008,qiu2008,panagia1973}.

In Figure 2 of \citet{martin2008}, MIR emission is also seen at the position of
the ``subcluster'' centered in sources 6-s, 7-s and 9-s (source mm2 in their
work and in \citealt{beuther2003}). There is no detected cm
counterpart at that position, suggesting that either none of the sources
have ignited or the ionized \hii\ region is still too small so the free-free
emission is confined and not detectable. Therefore these sources might be at an
even earlier stage of evolution as 13-s, the equivalent of a Class 0 low-mass
protostar, heating their circumstellar dust enough to be detected at MIR
wavelengths.

Source nr76 of \citet{martin2008} is located less than 1.5\arcsec\ from source
1-s. However, nr76 is detected in the J, H and K$_s$ bands and not in any of the
Spitzer bands, while 1-s is marginally detected at \rxun\ and is resolved at
\rxtres. The lack of emission in the Spitzer bands while being well detected in
the K$_s$ band, and the fact that is a relatively strong source at \rxtres,
make us believe that nr76 is not the counterpart of 1-s, but a separate
source, appearing nearby due to a projection effect. The case is similar also
for source nr71. It is located at $\sim1.5''$ from 20-s, and although nr71 shows
strong MIR emission, we only detect source 20-s at \rxtres\ and not at \rxun.
Therefore we believe that this is also a case of spatial projection.

This can be expected, since \mia\ is embedded in a cluster of Young Stellar
Objects (YSOs). \citet{martin2008} detects 116 NIR/MIR sources in a
\mbox{$\sim75''\times75''$} region centered on \mia, while \citet{qiu2008}
detect 46 YSOs ranging from Class 0 to Class \textsc{ii} protostars within a
radius of $\sim2$\,pc, in addition to the detection of over 800 NIR sources in
that region. The fact that there are already low-mass stars in the outskirts of
\mia\ while high-mass stars are still forming in it  supports the hypothesis
that the low-mass stars form before their high-mass counterparts (e.g.,
\citealt{kumar2006}).

Also in the Northern (proto)cluster there are identified NIR and MIR
counterparts.
\citet{qiu2008} found 5-band Spitzer/IRAC and 2MASS $H$ and $K_s$ emission
towards 4-n, as well as Spitzer/IRAC $3.6\mu m, 4.5\mu m$, $5.8\mu m$ and
$8.0\mu m$ emission towards 1/2/3-n. Source 4-n is likely at an early stage of
evolution, however it is likely the most evolved source in the Northern
(proto)cluster. It is not detected at cm wavelengths, therefore it has either
not yet ignited its protostar or the ionized region is still too small and the
free-free emission is trapped and thus not yet detectable. The other NIR/MIR
detection in the (proto)cluster is towards sources 1/2/3-n, and their NIR/MIR
flux is lower than that of 4-n, while their respective \rxun\ masses are
similar. This would indicate that 4-n is in a more advanced evolutionary stage
than 1/2/3-n.

With a luminosity of $\sim10^4\, \mathrm{L}_{\odot}$, \mia\ is a high-mass
star-forming region, but the (proto)stellar content in its core is unknown
because of high obscuration. Since it shows cm emission, most likely it
already has a (proto)stellar component, ionizing an Ultra- or even a
Hypercompact \hii\ region. While it is not possible to directly measure the
masses of the (proto)stars with mm data (see Sec.
\ref{sec-res-cont}), it is possible to estimate the
masses of the circumstellar structure.
Our shortest baseline of 20m at the given distance
of $2.2$\,kpc corresponds to a spatial scale of $\sim38000$\,AU at \rxun,
therefore any structure larger than this was not detected.
Then the low masses that we calculate from the \rxun\ data can be attributed to
a circumstellar structure, while the surrounding envelope likely contributes to
the larger masses derived from the \rxtres\ data.

\begin{figure}[h]
	\centering
	\includegraphics[angle=-90, width=0.5\textwidth]{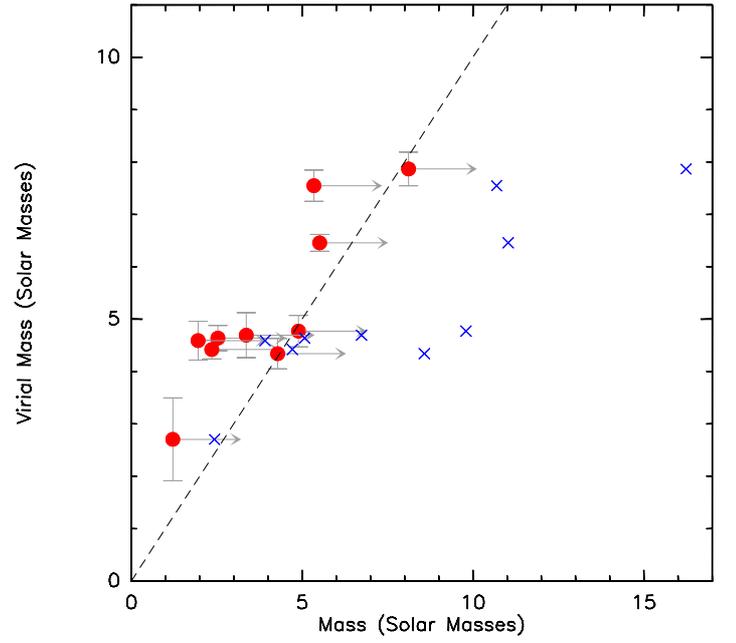}
	\caption{Comparison between the virial masses and gas masses of the
resolved cores in \mia. Virial masses are not affected by the interferometric
flux filtering, which do affect the gas masses. The red dots are the actual
values obtained, while the blue crosses represent the relationship if we take
into account $\sim10\%$ of the flux filtered out by the interferometer. This is
a hint of the care that must be taken when interpreting interferometric flux
values.}
	\label{fig-virial}
\end{figure}

\subsection{Virial and Jeans analysis}

The large uncertainty that the flux filtering introduces in the mass
calculations can be seen when comparing the gas masses with the virial masses.
The former are calculated from the interferometric continuum emission that is
affected by the flux filtering, while the latter are calculated using the
combined interferometric and single-dish \hdco\ data, with the short-spacings
correcting the effect introduced by the flux filtering.
The virial and gas masses for the resolved cores in \mia\ are compared in Figure
\ref{fig-virial}. At first glance, the cores do not appear to be collapsing,
however the gas masses derived from the continuum are lower than the virial
masses because of the missing flux discussed previously.
If we shift the gas masses higher by a factor of 2, taking into account only
$\sim10\%$ of the missing flux, we see that within the uncertainty on the virial
masses now all the resolved cores are likely collapsing.

The relative distances between neighboring cores within each (proto)cluster
range several thousands of AU.
These are similar and below the (proto)cluster's Jeans length
\mbox{$\lambda_J\sim 25000\,{\rm AU}$}, calculated from the equation

\begin{equation}
\label{eq_jeans-length}
	\lambda_J=0.19{\rm pc\left(\frac{T}{10
K}\right)^{1/2}\left(\frac{\nhdos}{10^4\,cm^{-3}}\right)^{-1/2}}
\end{equation}

\noindent for an average $\nhdos\sim10^5\,{\rm cm^{-3}}$ and an average $\tkin\sim40$\,K for the whole large-scale clumps
(see Sec. \ref{sec-lvg-modeling}; \citealt{stahler2005}). Our resolution of
$\sim1''$ at \rxun\ corresponds to a spatial scale of $\sim2200$\,AU, which
clearly resolves the Jeans length of the region. We are thus resolving the
fragmentation of the clump-scale (proto)clusters far below their Jeans
length.
This cannot be said in the case of the cores, for which our resolution is above
the Jeans length of the individual cores, in the $700 - 1200$\,AU range.

We are resolving the Jeans length of the clumps by one order of magnitude,
almost reaching the Jeans length of the individual cores, therefore we can
safely assume that we are mapping the direct progenitors of single stars or
multiple systems at most (similar to the Trapezium in Orion, see e.g.,
\citealt{rodon2008}). From the \hdco\ data, we see that there is only
a small signature of velocity dispersion (see Fig. \ref{fig-h2co-1moment}),
which would imply that the (proto)clusters are in a stage of weak dynamical
evolution.

\subsection{Considerations}
\label{sec-caveats}

\subsubsection{Masses from (sub)mm interferometry}
All interferometer observations underestimate the masses detected because of the incomplete sampling of the uv-plane. 
However, this filtering is not necessarily constant throughout the observed region. 
For instance, the single-dish observations map the cores as well as their envelope. This gas and dust reservoir is not necessarily gravitationally bound to the cores in its totality, thus only a fraction of it will be accreted by the cores during their evolution. Theory and observations cannot yet estimate how much this fraction is, and in addition, we are not sensitive to the mass that is already in (proto)stars or in small, optically thick cores. Therefore, the underestimation factor we have calculated for the masses is an upper-limit value, obtained in the simple approximation that all the mass detected with single-dish observations is uniformly distributed and contributes to each core in the same way.

Furthermore, the cores might not draw uniformly from the envelope during their accretion phase. 
The competitive accretion model (e.g., \citealt{bonnell2004,bonnell2007,peters2010}), for example, tells that the most massive cores will accrete more from the envelope than lower-mass cores, and that it cannot be pointed out to which (proto)star(s) some parts of the envelope are going to contribute.
It also claims that there maybe is the problem that the mass reservoir of each (proto)star is not a well defined concept. This has then to be added to the problem of the spatial filtering from the interferometer.

All considered, we believe that the core masses we have determined are the best values that can be obtained with the currently available instruments and techniques. New studies with ALMA will certainly improve significantly in this field.

\subsubsection{The nature of the CMF}

In an ideal case, the CMF should be calculated with only the pre-stellar (starless) cores because if there is already a protostar, the gas mass accreted onto it is not negligible when compared with its initial core mass.
However, in the high-mass case the starless phase is very short lived if existing at all (see e.g. \citealt{motte2007}). Also, some of the MSF theories propose that non-prestellar cores would not affect the resulting CMF (\citealt{clark2008}, see Sec. \ref{sec-disc-cmf} for more details).
 
We can assume a relationship of the mass function (MF) of a given region to its CMF as $CMF = f(M)*MF$, where $f(M)$ is the fraction we are able to observe.  If $f(M) = const.$, then our data are a fair representation of the CMF. If it increases with mass, for example because a larger fraction is resolved out for higher-mass cores, then the slope of the MF we measure is steeper than the true CMF. If it instead decreases with Mass, because a larger fraction is resolved out for low-mass cores,  the slope measured will be shallower than the true CMF. The shape of this $f(M)$ could be estimated by synthetically ``observing'' model CMFs. However, this is out of the scope of this paper..

Even without flux losses, unresolved substructure and projection effects complicate the picture even more.

Taking into account all these caveats, we consider that using a protostellar CMF as the one derived in this work is the best approach we can currently follow for a MSF region, though care has to be taken when comparing it with the IMF. ALMA with its much better uv-coverage and means to measure the complete flux will significantly improve our technical capabilities to study the CMF in detail.

\subsection{The Core Mass Function of \mia}
\label{sec-disc-cmf}

Most of the studies that trace core scales ($\sim0.01 $\,pc) are on low-mass star forming regions, since the distances involved are of a few hundred parsecs and therefore it is possible to recover such spatial scales with single-dish telescopes. Some examples are \citet{motte1998} and \citet{johnstone2000}, who from single-dish sub-mm observations of the $ \rho $ Ophiuchi region the authors derive in each case a mass function that resembles the IMF.
A similar result is obtained by \citet{konyves2010} with Herschel observations (i.e., in the far-IR spectrum) of the Aquila rift (see also \citealt{andre2010}).
Going up to a few solar-like masses, there are for example the works of \citet{johnstone2006} and \citet{nutter2007} on Orion. In this case, also from single-dish sub-mm observations, they recover a mass function similar to the IMF.

The mass functions of MSF regions have indeed been derived in many previous studies, however most of them with single-dish telescopes that due to the distances involved of more than $ \sim2 $\,kpc, can only recover scales on the order of $ 0.1 $\,pc. According to the criteria we follow that correspond to clump scales, i.e., structures that may form (proto)clusters and not just a single star or a small multiple system. The results show mass spectra with indices $ \beta $ between $ \sim-1.2 $ and $ \sim-3 $ (e.g. \citealt{kramer1998, kerton2001, tothill2002, shirley2003, mookerjea2004, reid2005, reid2006, beltran2006, munioz2007, ikeda2011})

The case of \mia\ is, to our knowledge, the only one where a high-mass star forming region that could be mapped down to core scales presented enough fragmentation to derive a CMF. 
For a MSF region, a mass function at similar spatial scales was derived by \citet{reid2005} for NGC~7538. Their smaller condensations have an average size of $ \sim0.06 $\,pc, in-between clumps and cores. In fact, they derive a mass function with a power-law index $ \beta=-2.0\pm0.3 $, which despite being comparable with the stellar IMF, is also halfway between the clump mass function value of $ -1.6 $ and the CMF value of $ -2.3 $ found in low-mass star forming regions and in this work for a MSF region.

Remarkably, \mia\ is a very rare object for showing 
more than 10 cores when resolved down to a
spatial scale of several thousand AU. Similar high-spatial resolution (sub)mm
observations of MSF regions typically resolve only a few cores. For example,
interferometric PdBI and SMA mm mappings of the MSF regions W3\,IRS 5,
IRAS\,06058+2138, IRAS\,06061+2151, IRAS\,05345+3157 and AFGL961 resolve
cores down to spatial scales between $\sim750$ and $\sim5000$\,AU, recovering in
all cases fewer than 10 cores in each region
(\citealt{fontani2009,williams2009,rodon2008,rodon2009phd}).
These regions are similar in distance and luminosity to \mia, and although
the gas mass varies, even the regions more massive than \mia\ show much less
fragmentation at similar spatial scales.
More recently, there is the work of \citet{bontemps2010} in the Cygnus~X region. They resolve 5 clumps down to spatial scales similar to the ones we obtained, recovering 23 fragments in total, from which they estimate 9 are probable high-mass protostars. Although the total number of cores is similar to this work, each clump fragments to `only' 7 cores or less.

It is unclear why \mia\ is highly fragmented.
It could be an effect of being at a different evolutionary stage, having a different chemical
composition or even through perturbation and/or a different formation processes.
Whichever the case, this apparent lack of fragmentation is preventing the
derivation of a CMF for other individual MSF regions, highlighting the
rarity of \mia.

In general, observations of low-mass star-forming regions suggest that the IMF of low-mass
stars is determined at early evolutionary stages (e.g., \citealt{testi1998,
motte1998,motte2001,alves2007}; \citealt{konyves2010}; see varying suggestions e.g.,
\citealt{goodwin2009}). This is not so clear in the case of MSF regions, because dynamical processes like
competitive accretion and merging of less massive protostars can shape the IMF
at later evolutionary stages (e.g.,
\citealt{bonnell2004,bonnell2007,bonnell2006}).

A similarity between the CMF of a massive star forming region and the IMF would suggest that the structure
and conditions within the molecular cloud could determine the IMF. This is essentially what was found by \citet{chabrier2010} in a statistical study of the relationship between the CMF and the IMF. They find that both functions are correlated, and they argue that the IMF would then be defined in the early stages of evolution, being only slightly modified by environmental effects.
The IMF and the CMF would also be similar if the relationship between the cores and the stars forming from them is
one-to-one or nearly one-to-one. This kind of relationship is supported by
theoretical models explaining the shape of the high-mass end of the IMF (e.g.,
\citealt{scalo1998a,padoan2002}), and by the apparently constant star formation
efficiency suggested both by theory and observations
\citep{matzner2000,alves2007}.

Also, in the \citet{clark2008} picture of competitive accretion, the shape of
the IMF is independent of when competitive accretion is halted.
This would suggest that the IMF is set early on in the evolution of the clump,
and therefore having non-prestellar cores at the moment of deriving a CMF would
not affect the final results.

Taking into account the caveats discussed in Section \ref{sec-caveats}, our derived power-law index $\beta$ is similar to the high-mass end of the Salpeter IMF, and one would then be tempted to do a direct comparison between them. However, the CMF we derive is not completely pre-stellar, since we cannot confidently extract all the (proto)stellar sources. This means that our CMF is ``contaminated'' with sources that could already be considered when calculating the local IMF in the region. Thus we cannot draw solid conclusions about the likeness of the CMF of MSF regions and the IMF until the former cannot be derived with just pre-stellar cores.

Even then, caution must be taken when doing a direct comparison between the CMF
and the IMF. Numerical simulations show that although the overall shape of the
IMF in the low-mass to intermediate-mass regime is robust against different core
evolution scenarios, further turbulent fragmentation of the cores may change the
high-mass slope of the IMF \citep{swift2008}. 

\section{Conclusions}

We resolve the two clumps of the MSF region \mia\ into 26 cores at
\rxun, with a spatial resolution of $\sim 2200$\,AU. This resolves the Jeans
length of the clumps, and the relative distances between the cores are similar
or smaller than the Jeans length corresponding to individual cores.
Also, the cores show only a marginal signature of velocity dispersion,
implying that the (proto)clusters are not in a strong dynamical evolution.

The approach of calculating temperatures using \hdco\ lines might not be suitable for studies at high-spatial resolution and density like this one. We find that the \hdco\ and the continuum emission only correlate for the brightest source in each (proto)cluster and not for the fainter sources, therefore preventing the derivation of the temperature of each single core. However, it allows to derive a temperature structure for the (proto)clusters as a whole.

The temperature structure of \mia\ was determined from its \hdco\ and \cian\
emission allowing the estimation of a CMF. Taking into account the
arbitrariness of the mass binning when deriving a mass function, we found a CMF
index $\beta=-2.3\pm0.2$ for core masses above $\sim 1.4$\,M\sun, confirming the previous results of \beuthert\ with increased confidence levels.

It was not possible to determine with enough confidence the evolutionary stage of all the sources detected (i.e, whether they are pre-, proto-, or stellar sources). Because of this, the CMF we derive is not a fully pre-stellar one, and therefore it cannot be considered as a true predecessor of the stellar IMF, but a step in between a pre-stellar CMF and the IMF. An additional important caveat is the difficulty to combine continuum single-dish and interferometer data because of the vastly different bandpasses. This induces missing flux problems severely affecting the mass estimates.

To the present day, mapping of the CMF has been done mostly in low-mass star forming regions. We have shown and discussed the caveats that involve such a study in MSF regions, which in turn explains the rarity of such studies so far.
Mapping a MSF region down to $ \sim0.01 $\,pc scales requires an angular resolution on the order of one arc-second for the closest regions, and tenths of arc-second if we go to larger distances. At (sub)mm wavelengths that can only be achieved with interferometry. The interferometers operating at this moment normally achieve $ \sim1'' $ resolution, but to go below it is not that common. Sensitivity can become also an issue, and interferometers naturally filter out a large part of the incoming flux. When fully operational, ALMA will improve significantly on these observational limitations. Despite the difficulties and assumptions that have to be made due to these limitations, the study of the Core Mass Function of high-mass star forming regions has great usefulness for the study of the origin of the IMF. It is already seen that for low-mass stars, the IMF and the CMF are indistinguishable, suggesting the kind of relationship between them. That is yet unknown for high-mass stars and therefore it needs to be addressed, both theoretically and observationally.

\begin{acknowledgements}
      J.A.R. \& H.B. acknowledge support by the \emph{Deut\-sche
For\-schungs\-ge\-mein\-schaft, DFG\/} project number BE~2578. J.A.R. also
acknowledges support from the \emph{International Max-Planck Research School for
Astronomy and Cosmic Physics} at the University of Heidelberg.
\end{acknowledgements}

\def\aj{AJ}%
\def\araa{ARA\&A}%
\def\apj{ApJ}%
\def\apjl{ApJ}%
\def\apjs{ApJS}%
\def\ao{Appl.~Opt.}%
\def\apss{Ap\&SS}%
\def\aap{A\&A}%
\def\aapr{A\&A~Rev.}%
\def\aaps{A\&AS}%
\def\azh{AZh}%
\def\baas{BAAS}%
\def\jrasc{JRASC}%
\def\memras{MmRAS}%
\def\mnras{MNRAS}%
\def\pra{Phys.~Rev.~A}%
\def\prb{Phys.~Rev.~B}%
\def\prc{Phys.~Rev.~C}%
\def\prd{Phys.~Rev.~D}%
\def\pre{Phys.~Rev.~E}%
\def\prl{Phys.~Rev.~Lett.}%
\def\pasp{PASP}%
\def\pasj{PASJ}%
\def\qjras{QJRAS}%
\def\skytel{S\&T}%
\def\solphys{Sol.~Phys.}%
\def\sovast{Soviet~Ast.}%
\def\ssr{Space~Sci.~Rev.}%
\def\zap{ZAp}%
\def\nat{Nature}%
\def\iaucirc{IAU~Circ.}%
\def\aplett{Astrophys.~Lett.}%
\def\apspr{Astrophys.~Space~Phys.~Res.}%
\def\bain{Bull.~Astron.~Inst.~Netherlands}%
\def\fcp{Fund.~Cosmic~Phys.}%
\def\gca{Geochim.~Cosmochim.~Acta}%
\def\grl{Geophys.~Res.~Lett.}%
\def\jcp{J.~Chem.~Phys.}%
\def\jgr{J.~Geophys.~Res.}%
\def\jqsrt{J.~Quant.~Spec.~Radiat.~Transf.}%
\def\memsai{Mem.~Soc.~Astron.~Italiana}%
\def\nphysa{Nucl.~Phys.~A}%
\def\physrep{Phys.~Rep.}%
\def\physscr{Phys.~Scr}%
\def\planss{Planet.~Space~Sci.}%
\def\procspie{Proc.~SPIE}%
\let\astap=\aap
\let\apjlett=\apjl
\let\apjsupp=\apjs
\let\applopt=\ao


\begin{thebibliography}{82}
\expandafter\ifx\csname natexlab\endcsname\relax\def\natexlab#1{#1}\fi

\bibitem[{{Alves} {et~al.}(2007){Alves}, {Lombardi}, \& {Lada}}]{alves2007}
{Alves}, J., {Lombardi}, M., \& {Lada}, C.~J. 2007, \aap, 462, L17

\bibitem[{{Anderson} {et~al.}(2010){Anderson}, {Zavagno}, {Rod{\'o}n},
  {Russeil}, {Abergel}, {Ade}, {Andr{\'e}}, {Arab}, {Baluteau}, {Bernard},
  {Blagrave}, {Bontemps}, {Boulanger}, {Cohen}, {Compi{\`e}gne}, {Cox},
  {Dartois}, {Davis}, {Emery}, {Fulton}, {Gry}, {Habart}, {Huang}, {Joblin},
  {Jones}, {Kirk}, {Lagache}, {Lim}, {Madden}, {Makiwa}, {Martin},
  {Miville-Desch{\^e}nes}, {Molinari}, {Moseley}, {Motte}, {Naylor}, {Okumura},
  {Pinheiro Gon{\c c}alves}, {Polehampton}, {Saraceno}, {Sauvage}, {Sidher},
  {Spencer}, {Swinyard}, {Ward-Thompson}, \& {White}}]{anderson2010}
{Anderson}, L.~D., {Zavagno}, A., {Rod{\'o}n}, J.~A., {et~al.} 2010, \aap, 518,
  L99+

\bibitem[{{Andr{\'e}} {et~al.}(2010){Andr{\'e}}, {Men'shchikov}, {Bontemps},
  {K{\"o}nyves}, {Motte}, {Schneider}, {Didelon}, {Minier}, {Saraceno},
  {Ward-Thompson}, {di Francesco}, {White}, {Molinari}, {Testi}, {Abergel},
  {Griffin}, {Henning}, {Royer}, {Mer{\'{\i}}n}, {Vavrek}, {Attard},
  {Arzoumanian}, {Wilson}, {Ade}, {Aussel}, {Baluteau}, {Benedettini},
  {Bernard}, {Blommaert}, {Cambr{\'e}sy}, {Cox}, {di Giorgio}, {Hargrave},
  {Hennemann}, {Huang}, {Kirk}, {Krause}, {Launhardt}, {Leeks}, {Le Pennec},
  {Li}, {Martin}, {Maury}, {Olofsson}, {Omont}, {Peretto}, {Pezzuto}, {Prusti},
  {Roussel}, {Russeil}, {Sauvage}, {Sibthorpe}, {Sicilia-Aguilar}, {Spinoglio},
  {Waelkens}, {Woodcraft}, \& {Zavagno}}]{andre2010}
{Andr{\'e}}, P., {Men'shchikov}, A., {Bontemps}, S., {et~al.} 2010, \aap, 518,
  L102+

\bibitem[{{Beltr{\'a}n} {et~al.}(2006){Beltr{\'a}n}, {Brand}, {Cesaroni},
  {Fontani}, {Pezzuto}, {Testi}, \& {Molinari}}]{beltran2006}
{Beltr{\'a}n}, M.~T., {Brand}, J., {Cesaroni}, R., {et~al.} 2006, \aap, 447,
  221

\bibitem[{{Beuther} {et~al.}(2002{\natexlab{a}}){Beuther}, {Kerp}, {Preibisch},
  {Stanke}, \& {Schilke}}]{beuther2002e}
{Beuther}, H., {Kerp}, J., {Preibisch}, T., {Stanke}, T., \& {Schilke}, P.
  2002{\natexlab{a}}, \aap, 395, 169

\bibitem[{{Beuther} \& {Schilke}(2004)}]{beuther2004c}
{Beuther}, H. \& {Schilke}, P. 2004, Science, 303, 1167

\bibitem[{{Beuther} {et~al.}(2002{\natexlab{b}}){Beuther}, {Schilke}, {Menten},
  {Motte}, {Sridharan}, \& {Wyrowski}}]{beuther2002a}
{Beuther}, H., {Schilke}, P., {Menten}, K.~M., {et~al.} 2002{\natexlab{b}},
  \apj, 566, 945

\bibitem[{{Beuther} {et~al.}(2005){Beuther}, {Schilke}, {Menten}, {Motte},
  {Sridharan}, \& {Wyrowski}}]{beuther2002erratum}
{Beuther}, H., {Schilke}, P., {Menten}, K.~M., {et~al.} 2005, \apj, 633, 535

\bibitem[{{Beuther} {et~al.}(2002{\natexlab{c}}){Beuther}, {Schilke},
  {Sridharan}, {Menten}, {Walmsley}, \& {Wyrowski}}]{beuther2002b}
{Beuther}, H., {Schilke}, P., {Sridharan}, T.~K., {et~al.} 2002{\natexlab{c}},
  \aap, 383, 892

\bibitem[{{Beuther} {et~al.}(2003){Beuther}, {Schilke}, \&
  {Stanke}}]{beuther2003}
{Beuther}, H., {Schilke}, P., \& {Stanke}, T. 2003, \aap, 408, 601

\bibitem[{{Beuther} {et~al.}(2002{\natexlab{d}}){Beuther}, {Walsh}, {Schilke},
  {Sridharan}, {Menten}, \& {Wyrowski}}]{beuther2002c}
{Beuther}, H., {Walsh}, A., {Schilke}, P., {et~al.} 2002{\natexlab{d}}, \aap,
  390, 289

\bibitem[{{Beuther} {et~al.}(2006){Beuther}, {Zhang}, {Sridharan}, {Lee}, \&
  {Zapata}}]{beuther2006c}
{Beuther}, H., {Zhang}, Q., {Sridharan}, T.~K., {Lee}, C.-F., \& {Zapata},
  L.~A. 2006, \aap, 454, 221

\bibitem[{{Bonnell} \& {Bate}(2006)}]{bonnell2006}
{Bonnell}, I.~A. \& {Bate}, M.~R. 2006, \mnras, 370, 488

\bibitem[{{Bonnell} {et~al.}(2007){Bonnell}, {Larson}, \&
  {Zinnecker}}]{bonnell2007}
{Bonnell}, I.~A., {Larson}, R.~B., \& {Zinnecker}, H. 2007, in Protostars and
  Planets V, ed. B.~{Reipurth}, D.~{Jewitt}, \& K.~{Keil}, 149--164

\bibitem[{{Bonnell} {et~al.}(2004){Bonnell}, {Vine}, \& {Bate}}]{bonnell2004}
{Bonnell}, I.~A., {Vine}, S.~G., \& {Bate}, M.~R. 2004, \mnras, 349, 735

\bibitem[{{Bontemps} {et~al.}(2010){Bontemps}, {Motte}, {Csengeri}, \&
  {Schneider}}]{bontemps2010}
{Bontemps}, S., {Motte}, F., {Csengeri}, T., \& {Schneider}, N. 2010, \aap,
  524, A18+

\bibitem[{{Carey} {et~al.}(1998){Carey}, {Clark}, {Egan}, {Price}, {Shipman},
  \& {Kuchar}}]{carey1998}
{Carey}, S.~J., {Clark}, F.~O., {Egan}, M.~P., {et~al.} 1998, \apj, 508, 721

\bibitem[{{Chabrier} \& {Hennebelle}(2010)}]{chabrier2010}
{Chabrier}, G. \& {Hennebelle}, P. 2010, \apjl, 725, L79

\bibitem[{{Clark} {et~al.}(2008){Clark}, {Bonnell}, \& {Klessen}}]{clark2008}
{Clark}, P.~C., {Bonnell}, I.~A., \& {Klessen}, R.~S. 2008, \mnras, 386, 3

\bibitem[{{Draine} {et~al.}(2007){Draine}, {Dale}, {Bendo}, {Gordon}, {Smith},
  {Armus}, {Engelbracht}, {Helou}, {Kennicutt}, {Li}, {Roussel}, {Walter},
  {Calzetti}, {Moustakas}, {Murphy}, {Rieke}, {Bot}, {Hollenbach}, {Sheth}, \&
  {Teplitz}}]{draine2007}
{Draine}, B.~T., {Dale}, D.~A., {Bendo}, G., {et~al.} 2007, \apj, 663, 866

\bibitem[{{Draper}(1998)}]{draper1998}
{Draper}, N.~R. 1998, {Applied regression analysis}, ed. N.~R. {Draper}

\bibitem[{{Fontani} {et~al.}(2009){Fontani}, {Zhang}, {Caselli}, \&
  {Bourke}}]{fontani2009}
{Fontani}, F., {Zhang}, Q., {Caselli}, P., \& {Bourke}, T.~L. 2009, \aap, 499,
  233

\bibitem[{{Frerking} {et~al.}(1982){Frerking}, {Langer}, \&
  {Wilson}}]{frerking1982}
{Frerking}, M.~A., {Langer}, W.~D., \& {Wilson}, R.~W. 1982, \apj, 262, 590

\bibitem[{{Goodwin} \& {Kouwenhoven}(2009)}]{goodwin2009}
{Goodwin}, S.~P. \& {Kouwenhoven}, M.~B.~N. 2009, \mnras, 397, L36

\bibitem[{{Goodwin} {et~al.}(2007){Goodwin}, {Kroupa}, {Goodman}, \&
  {Burkert}}]{goodwin2007}
{Goodwin}, S.~P., {Kroupa}, P., {Goodman}, A., \& {Burkert}, A. 2007, in
  Protostars and Planets V, ed. B.~{Reipurth}, D.~{Jewitt}, \& K.~{Keil},
  133--147

\bibitem[{{Hildebrand}(1983)}]{hildebrand1983}
{Hildebrand}, R.~H. 1983, \qjras, 24, 267

\bibitem[{{Ikeda} \& {Kitamura}(2011)}]{ikeda2011}
{Ikeda}, N. \& {Kitamura}, Y. 2011, \apj, 732, 101

\bibitem[{{Jansen} {et~al.}(1994){Jansen}, {van Dishoeck}, \&
  {Black}}]{jansen1994}
{Jansen}, D.~J., {van Dishoeck}, E.~F., \& {Black}, J.~H. 1994, \aap, 282, 605

\bibitem[{{Jansen} {et~al.}(1995){Jansen}, {van Dishoeck}, {Black}, {Spaans},
  \& {Sosin}}]{jansen1995}
{Jansen}, D.~J., {van Dishoeck}, E.~F., {Black}, J.~H., {Spaans}, M., \&
  {Sosin}, C. 1995, \aap, 302, 223

\bibitem[{{Johnstone} {et~al.}(2006){Johnstone}, {Matthews}, \&
  {Mitchell}}]{johnstone2006}
{Johnstone}, D., {Matthews}, H., \& {Mitchell}, G.~F. 2006, \apj, 639, 259

\bibitem[{{Johnstone} {et~al.}(2000){Johnstone}, {Wilson}, {Moriarty-Schieven},
  {Joncas}, {Smith}, {Gregersen}, \& {Fich}}]{johnstone2000}
{Johnstone}, D., {Wilson}, C.~D., {Moriarty-Schieven}, G., {et~al.} 2000, \apj,
  545, 327

\bibitem[{{Kerton} {et~al.}(2001){Kerton}, {Martin}, {Johnstone}, \&
  {Ballantyne}}]{kerton2001}
{Kerton}, C.~R., {Martin}, P.~G., {Johnstone}, D., \& {Ballantyne}, D.~R. 2001,
  \apj, 552, 601

\bibitem[{{Kirk} {et~al.}(2006){Kirk}, {Johnstone}, \& {Di
  Francesco}}]{kirk2006}
{Kirk}, H., {Johnstone}, D., \& {Di Francesco}, J. 2006, \apj, 646, 1009

\bibitem[{{K{\"o}nyves} {et~al.}(2010){K{\"o}nyves}, {Andr{\'e}},
  {Men'shchikov}, {Schneider}, {Arzoumanian}, {Bontemps}, {Attard}, {Motte},
  {Didelon}, {Maury}, {Abergel}, {Ali}, {Baluteau}, {Bernard}, {Cambr{\'e}sy},
  {Cox}, {di Francesco}, {di Giorgio}, {Griffin}, {Hargrave}, {Huang}, {Kirk},
  {Li}, {Martin}, {Minier}, {Molinari}, {Olofsson}, {Pezzuto}, {Russeil},
  {Roussel}, {Saraceno}, {Sauvage}, {Sibthorpe}, {Spinoglio}, {Testi},
  {Ward-Thompson}, {White}, {Wilson}, {Woodcraft}, \& {Zavagno}}]{konyves2010}
{K{\"o}nyves}, V., {Andr{\'e}}, P., {Men'shchikov}, A., {et~al.} 2010, \aap,
  518, L106+

\bibitem[{{Kramer} {et~al.}(1998){Kramer}, {Stutzki}, {Rohrig}, \&
  {Corneliussen}}]{kramer1998}
{Kramer}, C., {Stutzki}, J., {Rohrig}, R., \& {Corneliussen}, U. 1998, \aap,
  329, 249

\bibitem[{{Kroupa}(2002)}]{kroupa2002}
{Kroupa}, P. 2002, Science, 295, 82

\bibitem[{{Kumar} {et~al.}(2006){Kumar}, {Keto}, \& {Clerkin}}]{kumar2006}
{Kumar}, M.~S.~N., {Keto}, E., \& {Clerkin}, E. 2006, \aap, 449, 1033

\bibitem[{{Loren} \& {Mundy}(1984)}]{loren1984}
{Loren}, R.~B. \& {Mundy}, L.~G. 1984, \apj, 286, 232

\bibitem[{{Ma{\'{\i}}z Apell{\'a}niz} \&
  {{\'U}beda}(2005)}]{maiz-apellaniz2005}
{Ma{\'{\i}}z Apell{\'a}niz}, J. \& {{\'U}beda}, L. 2005, \apj, 629, 873

\bibitem[{{Mangum} \& {Wootten}(1993)}]{mangum1993}
{Mangum}, J.~G. \& {Wootten}, A. 1993, \apjs, 89, 123

\bibitem[{{Mart{\'{\i}}n-Hern{\'a}ndez}
  {et~al.}(2008){Mart{\'{\i}}n-Hern{\'a}ndez}, {Bik}, {Puga}, {N{\"u}rnberger},
  \& {Bronfman}}]{martin2008}
{Mart{\'{\i}}n-Hern{\'a}ndez}, N.~L., {Bik}, A., {Puga}, E., {N{\"u}rnberger},
  D.~E.~A., \& {Bronfman}, L. 2008, \aap, 489, 229

\bibitem[{{Matzner} \& {McKee}(2000)}]{matzner2000}
{Matzner}, C.~D. \& {McKee}, C.~F. 2000, \apj, 545, 364

\bibitem[{{Mookerjea} {et~al.}(2004){Mookerjea}, {Kramer}, {Nielbock}, \&
  {Nyman}}]{mookerjea2004}
{Mookerjea}, B., {Kramer}, C., {Nielbock}, M., \& {Nyman}, L.-{\AA}. 2004,
  \aap, 426, 119

\bibitem[{{Motte} {et~al.}(1998){Motte}, {Andre}, \& {Neri}}]{motte1998}
{Motte}, F., {Andre}, P., \& {Neri}, R. 1998, \aap, 336, 150

\bibitem[{{Motte} {et~al.}(2001){Motte}, {Andr{\'e}}, {Ward-Thompson}, \&
  {Bontemps}}]{motte2001}
{Motte}, F., {Andr{\'e}}, P., {Ward-Thompson}, D., \& {Bontemps}, S. 2001,
  \aap, 372, L41

\bibitem[{{Motte} {et~al.}(2007){Motte}, {Bontemps}, {Schilke}, {Schneider},
  {Menten}, \& {Brogui{\`e}re}}]{motte2007}
{Motte}, F., {Bontemps}, S., {Schilke}, P., {et~al.} 2007, \aap, 476, 1243

\bibitem[{{Motte} {et~al.}(2003){Motte}, {Schilke}, \& {Lis}}]{motte2003}
{Motte}, F., {Schilke}, P., \& {Lis}, D.~C. 2003, \apj, 582, 277

\bibitem[{{Mu{\~n}oz} {et~al.}(2007){Mu{\~n}oz}, {Mardones}, {Garay},
  {Rebolledo}, {Brooks}, \& {Bontemps}}]{munioz2007}
{Mu{\~n}oz}, D.~J., {Mardones}, D., {Garay}, G., {et~al.} 2007, \apj, 668, 906

\bibitem[{{M{\"u}hle} {et~al.}(2007){M{\"u}hle}, {Seaquist}, \&
  {Henkel}}]{muehle2007}
{M{\"u}hle}, S., {Seaquist}, E.~R., \& {Henkel}, C. 2007, \apj, 671, 1579

\bibitem[{{Mundy} {et~al.}(1987){Mundy}, {Evans}, {Snell}, \&
  {Goldsmith}}]{mundy1987}
{Mundy}, L.~G., {Evans}, II, N.~J., {Snell}, R.~L., \& {Goldsmith}, P.~F. 1987,
  \apj, 318, 392

\bibitem[{{Nutter} \& {Ward-Thompson}(2007)}]{nutter2007}
{Nutter}, D. \& {Ward-Thompson}, D. 2007, \mnras, 374, 1413

\bibitem[{{Ossenkopf} \& {Henning}(1994)}]{ossenkopf1994}
{Ossenkopf}, V. \& {Henning}, T. 1994, \aap, 291, 943

\bibitem[{{Padoan} \& {Nordlund}(2002)}]{padoan2002}
{Padoan}, P. \& {Nordlund}, {\AA}. 2002, \apj, 576, 870

\bibitem[{{Panagia}(1973)}]{panagia1973}
{Panagia}, N. 1973, \aj, 78, 929

\bibitem[{{Peters} {et~al.}(2010){Peters}, {Klessen}, {Mac Low}, \&
  {Banerjee}}]{peters2010}
{Peters}, T., {Klessen}, R.~S., {Mac Low}, M., \& {Banerjee}, R. 2010, \apj,
  725, 134

\bibitem[{{Qiu} {et~al.}(2008){Qiu}, {Zhang}, {Megeath}, {Gutermuth},
  {Beuther}, {Shepherd}, {Sridharan}, {Testi}, \& {De Pree}}]{qiu2008}
{Qiu}, K., {Zhang}, Q., {Megeath}, S.~T., {et~al.} 2008, \apj, 685, 1005

\bibitem[{{Rathborne} {et~al.}(2008){Rathborne}, {Jackson}, {Zhang}, \&
  {Simon}}]{rathborne2008}
{Rathborne}, J.~M., {Jackson}, J.~M., {Zhang}, Q., \& {Simon}, R. 2008, \apj,
  689, 1141

\bibitem[{{Reid} \& {Wilson}(2005)}]{reid2005}
{Reid}, M.~A. \& {Wilson}, C.~D. 2005, \apj, 625, 891

\bibitem[{{Reid} \& {Wilson}(2006)}]{reid2006}
{Reid}, M.~A. \& {Wilson}, C.~D. 2006, \apj, 644, 990

\bibitem[{{Ridge} \& {Moore}(2001)}]{ridge2001}
{Ridge}, N.~A. \& {Moore}, T.~J.~T. 2001, \aap, 378, 495

\bibitem[{{Rodmann} {et~al.}(2006){Rodmann}, {Henning}, {Chandler}, {Mundy}, \&
  {Wilner}}]{rodmann2006}
{Rodmann}, J., {Henning}, T., {Chandler}, C.~J., {Mundy}, L.~G., \& {Wilner},
  D.~J. 2006, \aap, 446, 211

\bibitem[{{Rod\'on}(2009)}]{rodon2009phd}
{Rod\'on}, J.~A. 2009, PhD thesis, Max-Planck-Institut f\"ur Astronomie,
  Heidelberg, Germany

\bibitem[{{Rod{\'o}n} {et~al.}(2008){Rod{\'o}n}, {Beuther}, {Megeath}, \& {van
  der Tak}}]{rodon2008}
{Rod{\'o}n}, J.~A., {Beuther}, H., {Megeath}, S.~T., \& {van der Tak}, F.~F.~S.
  2008, \aap, 490, 213

\bibitem[{{Rod{\'o}n} {et~al.}(2010){Rod{\'o}n}, {Zavagno}, {Baluteau},
  {Anderson}, {Polehampton}, {Abergel}, {Motte}, {Bontemps}, {Ade},
  {Andr{\'e}}, {Arab}, {Beichman}, {Bernard}, {Blagrave}, {Boulanger}, {Cohen},
  {Compiegne}, {Cox}, {Dartois}, {Davis}, {Emery}, {Fulton}, {Gry}, {Habart},
  {Halpern}, {Huang}, {Joblin}, {Jones}, {Kirk}, {Lagache}, {Lin}, {Madden},
  {Makiwa}, {Martin}, {Miville-Desch{\^e}nes}, {Molinari}, {Moseley}, {Naylor},
  {Okumura}, {Orieux}, {Pinheiro Gon{\c c}alves}, {Rodet}, {Russeil},
  {Saraceno}, {Sidher}, {Spencer}, {Swinyard}, {Ward-Thompson}, \&
  {White}}]{rodon2010}
{Rod{\'o}n}, J.~A., {Zavagno}, A., {Baluteau}, J., {et~al.} 2010, \aap, 518,
  L80+

\bibitem[{{Salpeter}(1955)}]{salpeter1955}
{Salpeter}, E.~E. 1955, \apj, 121, 161

\bibitem[{{Scalo} {et~al.}(1998){Scalo}, {Vazquez-Semadeni}, {Chappell}, \&
  {Passot}}]{scalo1998a}
{Scalo}, J., {Vazquez-Semadeni}, E., {Chappell}, D., \& {Passot}, T. 1998,
  \apj, 504, 835

\bibitem[{{Schilke} {et~al.}(1999){Schilke}, {Phillips}, \&
  {Mehringer}}]{schilke1999}
{Schilke}, P., {Phillips}, T.~G., \& {Mehringer}, D.~M. 1999, in The Physics
  and Chemistry of the Interstellar Medium, ed. V.~{Ossenkopf}, J.~{Stutzki},
  \& G.~{Winnewisser}, 330--+

\bibitem[{{Shirley} {et~al.}(2003){Shirley}, {Evans}, {Young}, {Knez}, \&
  {Jaffe}}]{shirley2003}
{Shirley}, Y.~L., {Evans}, N.~J., {Young}, K.~E., {Knez}, C., \& {Jaffe}, D.~T.
  2003, \apjs, 149, 375

\bibitem[{{Sridharan} {et~al.}(2002){Sridharan}, {Beuther}, {Schilke},
  {Menten}, \& {Wyrowski}}]{sridharan2002}
{Sridharan}, T.~K., {Beuther}, H., {Schilke}, P., {Menten}, K.~M., \&
  {Wyrowski}, F. 2002, \apj, 566, 931

\bibitem[{{Stahler} \& {Palla}(2005)}]{stahler2005}
{Stahler}, S.~W. \& {Palla}, F. 2005, {The Formation of Stars} (The Formation
  of Stars, by Steven W.~Stahler, Francesco Palla, pp.~865.~ISBN
  3-527-40559-3.~Wiley-VCH , January 2005.)

\bibitem[{{Stutzki} \& {Guesten}(1990)}]{stutzki1990}
{Stutzki}, J. \& {Guesten}, R. 1990, \apj, 356, 513

\bibitem[{{Swift} \& {Williams}(2008)}]{swift2008}
{Swift}, J.~J. \& {Williams}, J.~P. 2008, \apj, 679, 552

\bibitem[{{Testi} \& {Sargent}(1998)}]{testi1998}
{Testi}, L. \& {Sargent}, A.~I. 1998, \apjl, 508, L91

\bibitem[{{Tieftrunk} {et~al.}(1998){Tieftrunk}, {Gaume}, \&
  {Wilson}}]{tieftrunk1998}
{Tieftrunk}, A.~R., {Gaume}, R.~A., \& {Wilson}, T.~L. 1998, \aap, 340, 232

\bibitem[{{Tothill} {et~al.}(2002){Tothill}, {White}, {Matthews}, {McCutcheon},
  {McCaughrean}, \& {Kenworthy}}]{tothill2002}
{Tothill}, N.~F.~H., {White}, G.~J., {Matthews}, H.~E., {et~al.} 2002, \apj,
  580, 285

\bibitem[{{van der Tak} {et~al.}(2000){van der Tak}, {van Dishoeck}, \&
  {Caselli}}]{vandertak2000a}
{van der Tak}, F.~F.~S., {van Dishoeck}, E.~F., \& {Caselli}, P. 2000, \aap,
  361, 327

\bibitem[{{Watanabe} \& {Mitchell}(2008)}]{watanabe2008}
{Watanabe}, T. \& {Mitchell}, G.~F. 2008, \aj, 136, 1947

\bibitem[{{Williams} {et~al.}(2009){Williams}, {Mann}, {Beaumont}, {Swift},
  {Adams}, {Hora}, {Kassis}, {Lada}, \&
  {Rom{\'a}n-Z{\'u}{\~n}iga}}]{williams2009}
{Williams}, J.~P., {Mann}, R.~K., {Beaumont}, C.~N., {et~al.} 2009, \apj, 699,
  1300

\bibitem[{{Wilner} {et~al.}(1994){Wilner}, {Wright}, \&
  {Plambeck}}]{wilner1994}
{Wilner}, D.~J., {Wright}, M.~C.~H., \& {Plambeck}, R.~L. 1994, \apj, 422, 642

\bibitem[{{Wootten} {et~al.}(1978){Wootten}, {Evans}, {Snell}, \& {vanden
  Bout}}]{wootten1978}
{Wootten}, A., {Evans}, II, N.~J., {Snell}, R., \& {vanden Bout}, P. 1978,
  \apjl, 225, L143

\bibitem[{{Xu} {et~al.}(2009){Xu}, {Reid}, {Menten}, {Brunthaler}, {Zheng}, \&
  {Moscadelli}}]{xu2009}
{Xu}, Y., {Reid}, M.~J., {Menten}, K.~M., {et~al.} 2009, \apj, 693, 413

\bibitem[{{Zhang} {et~al.}(1998){Zhang}, {Ho}, \& {Ohashi}}]{zhang1998b}
{Zhang}, Q., {Ho}, P.~T.~P., \& {Ohashi}, N. 1998, \apj, 494, 636

\end{thebibliography}
\end{document}